\documentclass[final]{siamltex}

\usepackage{euscript,amsmath,amssymb,amsfonts,graphicx,cite}
\allowdisplaybreaks[1]

\usepackage{chngcntr}
\counterwithout{figure}{section}
\counterwithout{equation}{section}
\usepackage{xcolor}

\usepackage{epstopdf}

\newcommand{\e}{{\rm e}}
\renewcommand{\d}{{\rm d}}
\newcommand{\pd}{\partial}

\newcommand{\R}{{\mathbb R}}

\newcommand{\Z}{{\mathbb Z}}

\newcommand{\D}{\displaystyle}
\newcommand{\mc}{\mathcal }

\newcommand{\W}{{\mathcal W}}

\title{Threshold of front propagation in neural fields: \\ An interface dynamics approach}
\author{Gr\'egory Faye\thanks{CNRS, UMR 5219, Institut de Math\'{e}matiques de Toulouse, 31062 Toulouse Cedex, France ({\tt gregory.faye@math.univ-toulouse.fr})}\and Zachary P. Kilpatrick\thanks{Department of Applied Mathematics, University of Colorado, Boulder, Colorado 80309, USA ({\tt colorado.edu}). This author is supported by NSF grant (DMS-1615737).}}

\begin{document}

\maketitle

\begin{abstract} Neural field equations model population dynamics of large-scale networks of neurons. Wave propagation in neural fields is often studied by constructing traveling wave solutions in the wave coordinate frame. Nonequilibrium dynamics are more challenging to study, due to the nonlinearity and nonlocality of neural fields, whose interactions are described by the kernel of an integral term. Here, we leverage interface methods to describe the threshold of wave initiation away from equilibrium. In particular, we focus on traveling front initiation in an excitatory neural field. In a neural field with a Heaviside firing rate, neural activity can be described by the dynamics of the interfaces, where the neural activity is at the firing threshold. This allows us to derive conditions for the portion of the neural field that must be activated for traveling fronts to be initiated in a purely excitatory neural field. Explicit equations are possible for a single active (superthreshold) region, and special cases of multiple disconnected active regions. The dynamic spreading speed of the excited region can also be approximated asymptotically. We also discuss extensions to the problem of finding the critical spatiotemporal input needed to initiate waves.
\end{abstract}


\section{Introduction}
\label{intro}
Traveling waves are ubiquitous in nature, arising in a wide variety of biological processes, including epidemics~\cite{grenfell01}, actin polymerization~\cite{allard13}, and evolution~\cite{rouzine03}. These processes are usually modeled by nonlinear partial differential equations (PDE) that combine nonlinear local interactions and spatial dynamics like diffusion~\cite{murray01}. Such continuum equations can yield traveling wave solutions in closed form, so the effect of model parameters on wave dynamics can be quantified in detail. For instance, neural field models describe large-scale dynamics of nonlocally connected networks of neurons, and their constituent functions can be tuned to produce a multitude of spatiotemporal solutions~\cite{bressloff12}. Such results can be connected to coherent neural activity patterns observed in cortical slice and in vivo experiments~\cite{huang04,richardson05,huang10}. 

Large-scale neural activity imaged using voltage sensitive dye exhibits myriad forms of propagating neural activity in different regions of the brain~\cite{wu08,wang10}. For instance, sensory inputs can nucleate traveling waves in olfactory~\cite{delaney94} and visual cortices~\cite{han08}. Waves may propagate radially outward from the site of nucleation~\cite{ferezou06}, with constant direction as plane waves~\cite{xu07}, or rotationally as spiral waves~\cite{huang10}. Sufficiently large amplitude sensory stimuli can initiate traveling waves of neural activity, but the threshold for initiation is difficult to identify~\cite{sato12}. A recent study has shown that if two visual stimuli are presented sufficiently close together in time, only a single wave is generated~\cite{gao12}. This suggests there is an internal state-dependent threshold that shapes the time and stimulus-amplitude necessary for wave initiation. In this work, we analyze a neural field model to understand how such propagation thresholds can be defined in a large-scale network of neurons.

Neural field equations provide a tractable model of coherent neural activity, which can be used to relate features of a network to the activity patterns it generates~\cite{coombes05,bressloff12}. The building blocks of a neural field are excitatory neurons, which activate their neighbors, and inhibitory neurons, which inactivate their neighbors. Wilson and Cowan showed that a localized stimulus to an excitatory/inhibitory neural field can produce outward propagating traveling waves~\cite{wilson73}, and Amari constructed such solutions assuming a high gain firing rate function~\cite{amari77}. Following this seminal work, Ermentrout and McLeod used a continuation argument to prove the existence of traveling fronts in a purely excitatory neural field~\cite{ermentrout93}. Subsequent studies of neural fields have built on this work by incorporating propagation delays or spatial heterogeneity and by adding variables representing slow processes like adaptation~\cite{pinto01,hutt03,kilpatrick10,faye13,fang16}. A wide variety of spatiotemporal patterns emerge including Turing patterns~\cite{bressloff01b}, traveling pulses~\cite{pinto01,coombes05b,faye15}, breathers~\cite{folias04}, and self-sustained oscillations~\cite{troy07,kilpatrick10}. However, most previous work focuses on construction of solutions and local dynamics near equilibria, addressed via linear stability or perturbation theory~\cite{laing02,hutt03,coombes04}. Nonequilibrium dynamics are less tractable in these infinite-dimensional systems, and so there are few results exploring the outermost bounds of equilibrium solutions' basins of attraction.


In the present study, we characterize the basins of attraction of the stationary solutions of an excitatory neural field. We focus on a scalar neural field model that supports traveling front solutions~\cite{ermentrout93,pinto01}:
\begin{equation}
\begin{cases}
\pd_t u(x,t) = - u(x,t) + \int_{\R} w(x-y) H(u(y,t) - \kappa) \d y+ I(x,t),& \, t>0, \ \ x\in\R, \\
u(x,0) = u_0(x), \quad x\in \R,
\end{cases}\label{nfield}
\end{equation}
where $u(x,t)$ is the total synaptic input at location $x \in \R$ and time $t>0$ and $w(x-y)$ is a kernel describing synaptic connections from neurons at location $y$ to those at $x$. Our results can be extended to the case $x \in \R^2$, as we will show in a subsequent paper. We assume the kernel $w(x)$ is rotationally symmetric, $w(x) = w(|x|)$, decreasing in $|x|> 0$; positive, $w(x) > 0$; and has a bounded integral, $\int_{\R}w(x) \d x < \infty$. We also assume $\int_{\R}w(x) \d x =1$ for simplicity, but this is not essential to our findings. To calculate explicit results, we consider the exponential kernel~\cite{pinto01,bressloff01}
\begin{align}
w(x) = \frac{1}{2} \e^{-|x|}.  \label{exp}
\end{align}
Nonlinearity in Eq.~(\ref{nfield}) arises due to the Heaviside firing rate function~\cite{ermentrout93,coombes04}
\begin{align*}
H(u - \kappa) = \left\{ \begin{array}{cc} 1, & u \geq \kappa,  \\  0, & u < \kappa, \end{array} \right.
\end{align*}
allowing us to determine dynamics of Eq.~(\ref{nfield}) by the threshold crossings $u(x_j(t), t) = \kappa$, yielding interface equations~\cite{coombes11,coombes12}. Our analysis focuses on the case of Eq.~(\ref{nfield}) for which traveling fronts propagate outward, so active regions ($u(x,t) \geq \kappa$) invade inactive regions ($u (x,t) < \kappa$). As a consequence, throughout the manuscript we assume that $\kappa \in (0,1/2)$. We derive this condition explicitly in Section \ref{solutions}.
The central focus of our work is to examine how the long term dynamics of Eq.~(\ref{nfield}) are determined by the initial condition $u(x,t) = u_0(x)$. For simplicity, we restrict $0 \leq u_0(x) \leq 1$, $\forall x \in \R$. We also examine the impact of external inputs $I(x,t)$, determining how they shape the long term behavior of Eq.~(\ref{nfield}).

Several previous studies have shown that traveling front solutions to Eq.~(\ref{nfield}) can be constructed~\cite{ermentrout93,pinto01,bressloff01}. Importantly they coexist with the two stable homogeneous states, $u \equiv 0$ and $u \equiv 1$. Thus, some initial conditions $u_0(x)$ will decay ($u \to 0$), but others will propagate ($u \to 1$) as $t \to \infty$. Our work addresses the following question: What conditions on $u_0(x)$ and Eq.~(\ref{nfield}) determine $\lim_{t \to \infty} u(x,t)$? Note, in Section \ref{solutions}, we explicitly construct a family of unstable intermediate stationary solutions, including single bumps and periodic patterns. While it is tempting to consider these solutions separatrices between the quiescent state ($ u \equiv 0$) and the emergence of two counter-propagating fronts, this picture of the full dynamics of Eq.~(\ref{nfield}) is incomplete. One can easily construct initial conditions $u_0(x)$ whose long term dynamics cannot be resolved by simply examining properties of these intermediate solutions. To distinguish cases that lead to decay versus propagation, we project the neural field Eq.~(\ref{nfield}) dynamics to equations for the interfaces $x_j(t)$ where $u(x_j(t),t) = \kappa$.

The problem of identifying the boundary between wave propagation and extinction has been studied previously in nonlinear PDE models on $\R$~\cite{zlatovs06,du10}. For initial conditions given by an indicator function $\chi_{[-l,l]}$ (1 on $x \in [-l,l]$ and 0 otherwise), activity decays when $l< l_0$ and propagates when $l>l_0$ for a critical width $l_0$~\cite{kanel64,aronson75}. For precisely $l=l_0$, the dynamics evolves to a separatrix~\cite{zlatovs06}. These results were generalized to arbitrary one-parameter initial condition families, where a parameter $l$ scales the initial condition height~\cite{du10}, and there is a critical value $l_0$ separating propagation from extinction. Such one-parameter approaches break down in attempting to analyze Eq.~(\ref{nfield}), due to its nonlocality. In particular, multiple active regions (for which $u(x,t)\geq \kappa$) interact nontrivially due to the nonlocal network connectivity. We will discuss this in detail in Section 4, but we note that some intuition from nonlinear PDE models is applicable to the case when $u_0(x)$ is a unimodal function, and so notions of a separatrix determining long term behavior can be applied (Section 3).

Our paper proceeds as follows. In Section \ref{solutions}, we characterize entire solutions to Eq.~(\ref{nfield}), which are relevant for our analysis, noting there are (i) homogeneous states, $u(x,t) \equiv \bar{u} \in \{0,1\}$; (ii) traveling waves, $u(x,t) = U_f(x-ct)$; and (iii) a family of unstable stationary solutions, $u(x,t) = U_L(x)$ with period $L$. Next, in Section \ref{single}, we perform a detailed analysis of the nonequilibrium dynamics of Eq.~(\ref{nfield}) using interface equations, for the case where $u_0(x) >\kappa$ only on a single active region $x \in [x_1, x_2]$, which allows us to classify the threshold between propagation ($u \to 1$) and failure ($u \to 0$). Our reduced interface equations also allow us to calculate the timescale of the transient dynamics as they approach equilibrium. In addition, we discuss requirements on an external stimulus $I(x,t)$ necessary to activate a traveling wave. In Section \ref{multiple}, we derive interface equations for Eq.~(\ref{nfield}) for multiple ($N>1$) active regions $u_0(x)$ for $x \in (x_1,x_2) \cup \cdots \cup (x_{2N-1},x_{2N})$. Some explicit results are possible in the cases $N \to \infty$ and $N=2$, showing interactions between active regions impact the propagation threshold. Our analysis provides a tool for linking initial conditions of spatially-extended neural field equations away from equilibrium to their eventual equilibrium state.

\section{Entire solutions of the excitatory neural field}
\label{solutions}
We begin by examining the entire solutions of the neural field Eq.~(\ref{nfield}) for $I(x,t) \equiv 0$. By entire solutions, we mean solutions of Eq.~(\ref{nfield}) which are defined for all time $t>0$. They divide into two classes: stationary solutions and traveling wave solutions. More precisely, we will show that the excitatory neural field Eq.~(\ref{nfield}) supports the following stationary solutions:
{\em
\begin{itemize}
\item[(i)] the two homogeneous states $u=0$ and $u=1$, which are both locally stable;
\item[(ii)] an unstable symmetric one bump solution $U_b$;
\item[(iii)] a family of periodic solutions $U_L$ which are all unstable.
\end{itemize}}
We conjecture that there are no other stationary solutions to the neural field Eq.~(\ref{nfield}). Proving this non-existence result is beyond the scope of the present paper, so we only cover the cases of standing front solutions and symmetric 2-bump solutions; see Section~\ref{expex} and Fig.~\ref{fig2_pport} for an illustration using the exponential kernel, Eq.~(\ref{exp}).

Our analysis of traveling waves focuses on fronts connecting  $\lim_{x \to -\infty} u(x,t)=1$ to $\lim_{x \to +\infty} u(x,t) = 0$, as these solutions are crucial for the analysis of subsequent sections. Finally, it is important to stress that the homogeneous states are stable, attracting almost all initial conditions (Fig.~\ref{fig1_separatrix}A,B), whereas the other stationary states separate some initial conditions into those that propagate and those that decay (Fig.~\ref{fig1_separatrix}B). However, there are other initial conditions, particularly multimodal initial conditions (Fig.~\ref{fig1_separatrix}C) that cannot be characterized using local analysis. Our analysis in Sections \ref{single} and \ref{multiple} will emphasize the nonequilibrium dynamics away from entire solutions, exploring conditions necessary for attraction to one of the two homogeneous states. Most of the results in this section are well-known~\cite{amari77,ermentrout98,folias04,coombes05,bressloff12,avitabile15}, so we refrain from excessive detail.  However, to the best of our knowledge, the results on unstable periodic solutions to Eq.~(\ref{nfield}) are new, along with the non-existence of a symmetric $2$-bump solution of Section~\ref{nonex}.

\subsection{Homogeneous states}
\label{homstates}
\begin{figure}
\begin{center} \includegraphics[width=13cm]{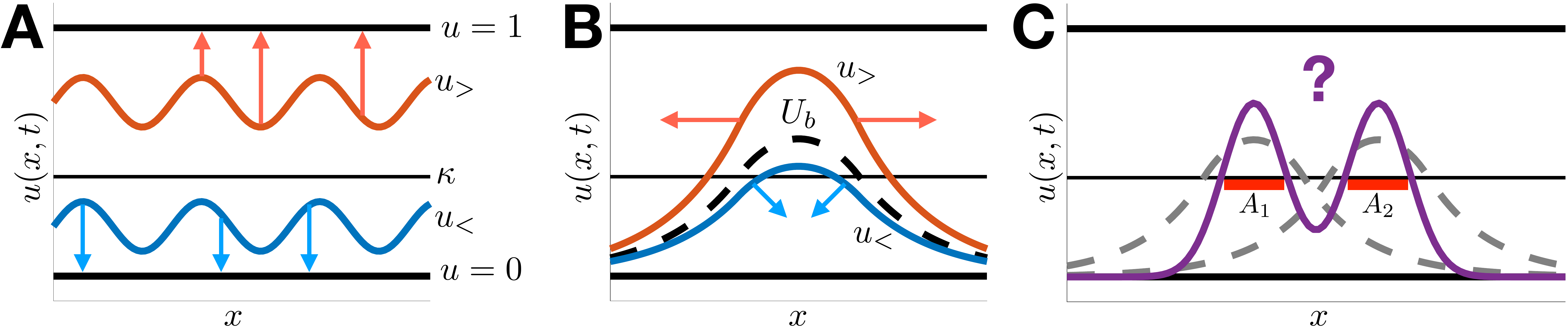} \end{center}
\vspace{-3mm}
\caption{Long term behavior of initial conditions $u_0(x)$ for Eq.~(\ref{nfield}) in 1D. (A) Entirely subthreshold (superthreshold) initial conditions decay (grow). If $u_0(x) < \kappa$, $\forall x$, then $u \to 0$ as $t \to \infty$ ($u_<$), whereas if $u_0(x) \geq \kappa$, $\forall x$, then $u \to 1$ as $t \to \infty$. (B) Initial conditions below (above) the unstable bump $U_b(x)$ decay (grow). If $u_0(x) < U_b(x)$, $\forall x$, then $u \to 0$ as $t \to \infty$ ($u_<$), whereas if $u_0(x)> U_b(x)$, $\forall x$, then $u \to 1$ as $t \to \infty$. (C) Characterization of $\lim_{t \to \infty}u(x,t)$ is less straightforward for multimodal initial conditions. Even though each active region ($A_1$ and $A_2$, where $u_0(x) \geq \kappa$) is narrower than the unstable bump $U_b(x)$, this initial condition could lead to propagation due to nonlocal interactions.}
\vspace{-5mm}
\label{fig1_separatrix}
\end{figure}

Spatially homogeneous solutions to Eq.~(\ref{nfield}) can be constructed by assuming the neural field is constant in space and time, so $u(x,t) \equiv \bar{u}$. This reduces Eq.~(\ref{nfield}) to $\bar{u} = \bar{w} H(\bar{u}- \kappa)$ where $\bar{w} = \int_{\R} w(x) \d x$, which we fix to be unity, $\bar{w}\equiv 1$, so $\bar{u} = H(\bar{u} - \kappa)$. Clearly, $\bar{u} = 0,1$ are the only solutions to this fixed point equation, and are both locally attractive. Consider Eq.~(\ref{nfield}) with $I \equiv 0$, where the initial condition satisfies $u_0(x) < \kappa$, $\forall x \in \R$, then $u(x,t) = u_0(x) \e^{-t}$ for all $t>0$, and $\lim_{t \to \infty} u(x,t) = 0$.
Similarly, for $u_0(x) \geq \kappa$, $\forall x \in \R$, then $u(x,t)=1-(1-u_0(x))\e^{-t}$ for all $t>0$, and $\lim_{t \to \infty} u(x,t) = 1$. In each case, we have pointwise convergence to one the two homogeneous states $\bar{u}=0,1$ (Fig. \ref{fig1_separatrix}A).

\subsection{Intermediate bump solution}\label{separatrix1d}

One special solution to the neural field Eq.~(\ref{nfield}) is a stationary bump, which is locally unstable. To construct the bump, we first assume a stationary solution, $u(x,t) = U_b(x)$, with a single active region $U_b(x) \geq \kappa$ for $x \in [x_1, x_2]$, which can be centered at $x=0$, so $x \in [-b,b]$. The solution then has the form
\begin{align}
U_b(x) = \int_{-b}^{b} w(x-y) \d y=W(x+b)-W(x-b), \label{statbump}
\end{align}
where we have defined the antiderivative of the weight kernel
\begin{align}
W(x) = \int_0^{x} w(y) \d y.  \label{wanti}
\end{align}
The threshold condition $U_b(\pm b) = W(2b)  = \kappa$ defines an implicit equation for the bump half-width $b$. Since $w(x)>0$ by assumption, $b \mapsto W(2b)$ is a strictly increasing function with $W(0)=0$ and $W_{\infty} : = \lim_{x \to \infty} W(x)=1/2$, so $\kappa = W(2b)$ has a unique solution
\[
b_0(\kappa)=W^{-1}(\kappa)/2
\]
for any $\kappa\in(0,1/2)$,  which can be computed by inverting $W$.

Local stability can be studied by tracking the evolution of small smooth perturbations via linearization of Eq.~(\ref{nfield}) around $U_b(x)$. To leading order, perturbations of the form  $u(x,t)=U_b(x)+\bar\psi(x,t)$ have dynamics given by the linear equation~\cite{amari77,folias04}
\begin{align*}
\frac{\partial \bar{\psi}(x,t)}{\pd t} = - \bar{\psi}(x,t) + \frac{1}{|U_b'(b)|} \left[ w(x+b) \bar{\psi}(-b,t) + w(x-b) \bar{\psi}(b,t) \right],
\end{align*}
assuming $\bar{\psi}(\pm b,t) \neq 0$. Other classes of perturbations do not contribute to instabilities. Assuming separable perturbations $\bar{\psi}(x,t) = \e^{\lambda t} \psi(x)$, we partition solutions into odd $\psi (-b) = -\psi (b)$ and even $\psi (b) = \psi (-b)$ symmetric functions~\cite{amari77,guo05}. For odd solutions, the eigenvalue
\[
\lambda_o = -1 + \left[ w(0) -w (2b) \right]/ \left[ w(0) - w(2b) \right] = 0.
\]
For even solutions, the associated eigenvalue
\[
\lambda_e = -1 + \left[ w(0)+w(2b) \right]/ \left[ w(0) - w(2b) \right] = 2w(2b) /\left[ w(0) - w(2b) \right] > 0,
\]
since $w(2b)$ is positive, so the bump solution is unstable.

\subsection{Periodic solutions}
\label{persolns}

The two homogeneous states ($u\equiv0$ and $u\equiv1$) and the intermediate bump solution $U_b$ are not the only possible stationary solutions of Eq.~(\ref{nfield}). Indeed, there exists a family of periodic solutions, parametrized by their period $L$, which can be explicitly constructed. Via translation symmetry, we can restrict our study to solutions with an active region centered at $x=0$. We denote by $U_L(x)$ an $L$-periodic solution of Eq.~(\ref{nfield}) which satisfies
\begin{equation}
\begin{cases}
U_L(x)> \kappa, & \text{ for } x \in (-b+nL,b+nL), \quad n \in \Z, \\
U_L(x)= \kappa & \text{ for }  x=\pm b+n L, \quad n \in \Z, \\
U_L(x)< \kappa, & \text{ elsewhere.}
\end{cases}
\label{periodic}
\end{equation}
We also impose that $2b<L$, otherwise $U_L(x)\geq \kappa$, $\forall x \in \R$, implying $u \equiv 1$ due to our previous analysis. Applying Eq.~\eqref{periodic} to Eq.~(\ref{nfield}), we find
\begin{align}
U_L(x)&= \sum_{n\in\Z} \int_{-b+nL}^{b+nL} w(x-y)\d y = \sum_{n\in\Z} \left( W(x+b+nL)-W(x-b+nL) \right).  \label{statper}
\end{align}
Applying any threshold condition, $U_L(\pm b + nL)=\kappa$, we obtain an implicit equation for $b$ given by
\begin{align}
\kappa = \sum_{n\in\Z} \left( W(2b+nL)-W(nL) \right):=\mathbf{W}_L(b),  \label{WL}
\end{align}
assuming $2b<L$. Let us remark, that $W(2b+nL)-W(nL)=2b w(\zeta_n)$ with $\zeta_n \in (nL,2b+nL)$ for all $n\in\Z$. Thus, for sufficiently localized $w$, we have that
\begin{equation*}
\lim_{L \to \infty} \sum_{n\in\Z, n \neq 0} \left( W(2b+nL)-W(nL) \right) = 2b \cdot \lim_{L \to \infty} \sum_{n\in\Z, n \neq 0} w(\zeta_n) = 0
\end{equation*}
so in the limit $L\rightarrow +\infty$, we recover the condition for a one bump solution, $\kappa =W(2b)$. Furthermore, by definition we have $\mathbf{W}_L(0)=0$ and for $L< \infty$, 
\begin{equation*}
\underset{b\rightarrow L/2}{\lim} ~\mathbf{W}_L(b)  = \int_{- \infty}^{\infty} w(x) \d x =1.
\end{equation*}
Also by construction,  $b\mapsto \mathbf{W}_L(b)$ is strictly increasing. As a conclusion, for any given $\kappa \in (0,1)$ and $L>0$,  the equation $\kappa = \mathbf{W}_L(b)$ has a  unique solution given by
\begin{equation*}
b_L(\kappa)=\mathbf{W}^{-1}_L(\kappa)\in\left( 0, L/2 \right).
\end{equation*}

Local stability of $U_L(x)$ can be determined by linearizing Eq.~(\ref{nfield}) using $u(x,t)=U_L(x)+\bar{\psi}(x,t)$ and expanding to first order in $\bar{\psi}$.
Decomposing solutions of the linearized equation as $\bar{\psi}(x,t)=\psi (x)e^{\lambda t}$, we have the eigenvalue problem
\begin{equation*}
(\lambda+1)\psi(x)=\int_{\R} w(x-y)H'(U_L(y)-\kappa)\psi(y)\d y.
\end{equation*}
Noting $H'(U_L(x)-\kappa) \cdot |U_L'(b)| =\sum_{n\in\Z}\left(\delta(x-b-nL)+\delta(x+b-nL)\right)$, we find
\begin{equation}
(\lambda+1)\psi(x)=\sum_{n\in\Z} \frac{\left(w(x-b-nL)\psi(b+nL)+w(x+b-nL)\psi(-b+nL) \right)}{|U'_L(b)|}.
\label{eigper}
\end{equation}
We first focus on odd and even $L$-periodic perturbations. For odd perturbations, $\psi(b+nL)=-\psi(-b+nL)=\psi_o\neq0$ for all $n\in\Z$. Plugging into Eq.~\eqref{eigper} and evaluating at $x=b$, we find the associated eigenvalue is
\begin{equation*}
\lambda_o=-1+\frac{1}{|U'_L(b)|}\sum_{n\in\Z} \left(w(nL)-w(2b+nL)\right)=0,
\end{equation*}
as Eq.~(\ref{statper}) implies
\begin{align}
|U'_L(b)|=\sum_{n\in\Z} \left(w(nL)-w(2b+nL)\right).  \label{uperprime}
\end{align}
The fact that $\lambda_o=0$ is due to the translation invariance of Eq.~(\ref{nfield}). Next, we consider even perturbations of the form $\psi(b+nL)=\psi(-b+nL)=\psi_e\neq0$ for all $n\in\Z$, leading to a corresponding eigenvalue
\begin{equation*}
\lambda_e=-1+\frac{1}{|U'_L(b)|}\sum_{n\in\Z} \left(w(nL)+w(2b+nL)\right)>0.
\end{equation*}
Indeed, we have used the fact
\begin{equation*}
\sum_{n\in\Z} \left(w(nL)+w(2b+nL)\right)-\sum_{n\in\Z} \left(w(nL)-w(2b+nL)\right)=2 \sum_{n\Z}w(2b+nL)>0.
\end{equation*}
Thus, periodic solutions $U_L(x)$ are always linearly unstable with respect to even perturbations. In fact, it is possible to check that even perturbations are the most unstable perturbations among the class of $L$-periodic perturbations. Indeed, for generic perturbations of the form
\begin{equation*}
\psi(x)=\psi_k e^{i k x \frac{2\pi}{L}}, \quad \psi_k \neq 0, \quad k\in\Z,
\end{equation*}
we obtain from Eq.~\ref{eigper} and evaluating at $x=b$ that
\begin{equation*}
\lambda_k=-1+\frac{1}{|U'_L(b)|}\sum_{n\in\Z}\left( w(nL)+w(2b+nL)e^{-2ikb\frac{2\pi}{L}}\right).
\end{equation*}
Taking the real part, we get
\begin{equation*}
\mathrm{Re}(\lambda_k)=-1+\frac{1}{|U'_L(b)|}\sum_{n\in\Z}\left( w(nL)+w(2b+nL)\cos\left(2kb\frac{2\pi}{L}\right)\right),
\end{equation*}
from which we deduce that $\mathrm{Re}(\lambda_k)\leq \mathrm{Re}(\lambda_0)$, $\forall k \in \Z$, and $\lambda_0$ precisely corresponds to even perturbations. As a consequence, even $L$-periodic perturbations are the most unstable $L$-periodic perturbations.

\subsection{Non-existence of other stationary solutions}\label{nonex}

Since stationary solutions with one bump as well as an infinite number of bumps exist in Eq.~(\ref{nfield}), one wonders if other forms of stationary solutions exist. While the general problem is difficult to address, we can rule out two specific cases. In particular, we can demonstrate there are no symmetric $2$-bump solutions or standing front solutions. \\
\vspace{-3mm}

\noindent
\textbf{Symmetric 2-bumps.} To construct a contradiction, we begin by assuming there exists a symmetric $2$-bump solution, with active regions ($U(x) \geq \kappa$) supported on $x \in [-b,-a]\cup[a,b]$ for $0<a<b$. The profile $U(x)$ of the $2$-bump solution satisfies
\begin{equation*}
U(x) = W(x+b)-W(x+a)+W(x-a)-W(x-b),
\end{equation*}
for all $x\in\R$. Enforcing the threshold conditions $\kappa=U(\pm a)=U(\pm b)$ then yields a system of two equations
\begin{align*}
\kappa &= W(a+b)-W(2a)-W(a-b),\\
\kappa &= W(2b)-W(a+b)+W(b-a).
\end{align*}
Subtracting the equations and noting $W(b-a)=-W(a-b)$, we find
\begin{equation*}
W(a+b)=\frac{W(2a)+W(2b)}{2}, \quad 0<a<b.
\end{equation*} 
Let us now note that $x\mapsto W(x)$ is strictly concave on $\R_+$ as $W''(x)=w'(x)<0$ for all $x>0$ by assumption. As a consequence, for any $0<a<b$, we have
\begin{equation*}
W(a+b)>\frac{W(2a)+W(2b)}{2},
\end{equation*}
which is a contradiction, so there are no symmetric $2$-bump solutions. \\
\vspace{-3mm}

\noindent
{\bf Standing front solutions.} Assume there exists a standing front with $U(x) \gtrless \kappa$ for $x \lessgtr a$. By translation symmetry of Eq.~(\ref{nfield}), we can set $a=0$ without loss of generality. In this case, $U(x) = \int_{0}^{\infty} w(x-y) \d y = W_{\infty} - W(x) $, so the threshold condition $\kappa = U(0)$ implies $\kappa = W_{\infty} = 1/2$, which is not true if $\kappa \in (0,1/2)$, and we have a contradiction. \\
\vspace{-3mm}

\subsection{An illustrative example: exponential weight kernel}
\label{expex}
\begin{figure}
\begin{center} \includegraphics[width=6cm]{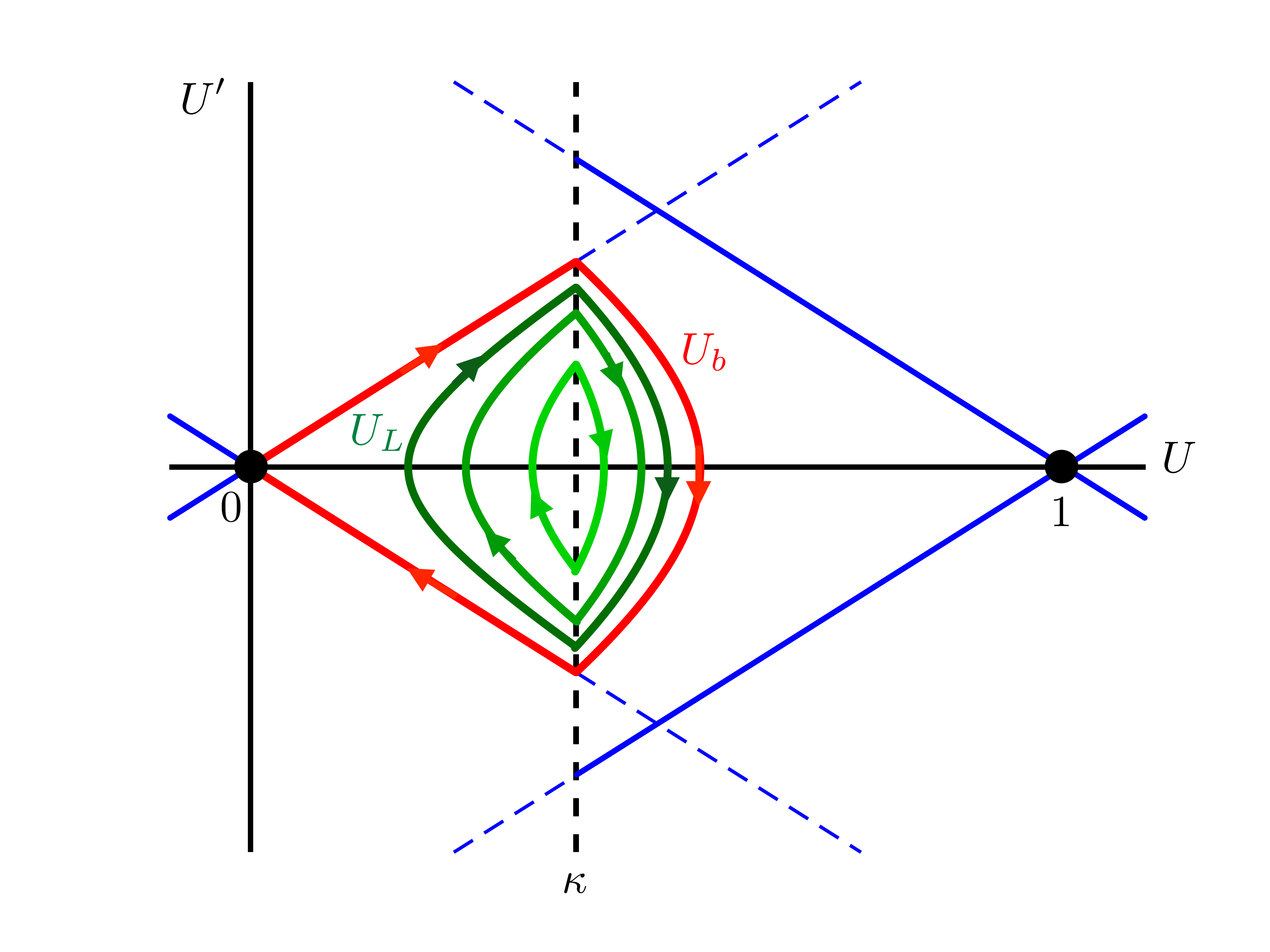} \end{center}
\vspace{-2mm}
\caption{Phase portrait of Eq.~(\ref{ODE}), describing stationary solutions of Eq.~(\ref{nfield}) with an exponential kernel, Eq.~(\ref{exp}) with $\kappa\in(0,1/2)$. Solid black and blue lines are nullclines of $U$ and $U'$, respectively. Homogeneous states $\bar{U}=0,1$ occur at their intersection. Homoclinic orbits arise about the point $(U,U')=(\kappa, 0)$, crossing the threshold $\kappa$ twice. The single bump $U_b$ (red trajectory) forms a separatrix, bounding all other nontrivial stationary solutions. There exists an infinite number of periodic solutions $U_L$ inside (e.g., green trajectories), whose orbits shrink as $L$ is decreased from infinity.}
\vspace{-5mm}
\label{fig2_pport}
\end{figure}

It is illustrative to consider the special case of the exponential weight kernel, Eq.~(\ref{exp}). In this case, the Fourier transform of $w$ is $\hat{w}(k)=1/(1+k^2)$ for $k\in\R$, so the convolution by $w$ corresponds to the operator $(\mathrm{I}-\partial_{xx})^{-1}$. As a consequence, any stationary solutions are solutions of the following piecewise-smooth second order differential equation, $U(x)-U''(x)=H(U(x)-\kappa)$, which can be written as
\begin{equation}
\begin{cases}
U'(x)&=V(x),\\
V'(x)&=U(x)-H(U(x)-\kappa).
\end{cases}\label{ODE}
\end{equation}
Eq.~(\ref{ODE}) is relatively easy to study as it reduces to two dynamical systems, depending on whether $U(x)\gtrless \kappa$. For $U(x)<\kappa$, we have $V'(x) = U(x)$, while for $U(x) \geq \kappa$, we find $V'(x) = U(x)-1$.
The complete phase portrait of Eq.~\eqref{ODE} is given in Fig.~\ref{fig2_pport} from which we recover the existence of a unique symmetric bump solution and a family of periodic solutions. The non-existence of $N$-bump solutions is then a trivial consequence of the phase portrait analysis and the monotonicity properties of the vector field associated to Eq.~\eqref{ODE}.

\subsection{Traveling fronts} 

To construct traveling wave solutions, we introduce the traveling wave coordinate $\xi = x - ct$, where $c$ denotes the wave speed, and set $u(x,t) = U_f(\xi)$. This will be a heteroclinic orbit that connects $\lim_{\xi \to - \infty} U_f(\xi) = 1$ to $\lim_{\xi \to + \infty} U_f(\xi) \to 0$. We fix the threshold crossing point at $\xi = 0$, $U_f(0) = \kappa$, so $U_f(\xi) \gtrless 0$ for $\xi \lessgtr 0$. These assumptions can be applied to Eq.~(\ref{nfield}), and the corresponding equation integrated to yield
\begin{align*}
U_f(\xi) = \e^{\xi/c} \left[ \kappa - \frac{1}{c} \int_0^{\xi} \e^{-y/c} (W_{\infty} - W(y)) \d y \right].
\end{align*}
Assuming $c>0$ (for $\kappa \in (0, 1/2)$) and requiring boundedness implies
\begin{align}
\kappa = \frac{1}{c} \int_0^{\infty} \e^{-y/c} (W_{\infty} - W(y)) \d y,  \label{frontthresh}
\end{align}
and so the traveling front solution will be of the form
\begin{align}
U_f(\xi) = \frac{1}{c} \int_0^{\infty} \e^{-y/c} (W_{\infty} - W(y+ \xi)) \d y.  \label{frontsol}
\end{align}
Eq.~(\ref{frontthresh}) relates the wavespeed $c$ to the threshold $\kappa$ and kernel $w(x)$, and can be rearranged along with integration by parts to yield a simpler implicit equation for $c$,
\begin{align}
\int_0^{\infty} \e^{-y/c} w(y) \d y = W_{\infty} - \kappa. \label{frontthresh2}
\end{align}
Since $W_{\infty} = 1/2$, Eq.~(\ref{frontthresh2}) will only have a solution with corresponding $c \in (0,\infty)$ if $\kappa \in (0,1/2)$, since the integral on the left hand side is positive and bounded from above by $W_{\infty}$. For the case of an exponential kernel, Eq.~(\ref{exp}), we have $c = \frac{1}{2 \kappa} \left[ 1 - 2 \kappa \right]$, defining a unique rightward traveling front solution for fixed $\kappa \in (0,1/2)$.

Local stability of the traveling front is determined by studying the evolution of perturbations $u(x,t)=U_f(\xi)+\bar{\psi}(\xi,t)$, in wave coordinates $\xi$, and linearizing~\cite{coombes04}
\begin{align*}
\frac{\pd \bar{\psi}(\xi,t)}{\pd t} =c \frac{\pd \bar{\psi}(\xi,t)}{\pd \xi} - \bar{\psi}(\xi,t) + \frac{1}{|U_f'(0)|} w(\xi) \bar{\psi}(0,t).
\end{align*}
Restricting to perturbations of the form $\bar{\psi}(\xi,t)=\bar\psi(\xi)e^{\lambda t}$ with $\psi(0) \neq 0$, we find the following eigenvalue problem
\begin{equation*}
\bar \psi(\xi) = \frac{e^{\frac{(1+\lambda)\xi}{c}}}{c|U_f'(0)|}\left(\int_{\xi}^{+\infty}e^{\frac{-(1+\lambda)y}{c}}w(y)\d y\right)\bar \psi(0).
\end{equation*}
Evaluating at $\xi=0$, we see $\lambda$ are zeros of the Evans function $\mathcal{E}(\lambda)$ defined~\cite{coombes04}
\begin{equation*}
\mathcal{E}(\lambda)=1-\frac{\mathcal{H}(\lambda)}{\mathcal{H}(0)}, \hspace{9mm} \mathcal{H}(\lambda):=\int_0^{+\infty}e^{\frac{-(1+\lambda)y}{c}}w(y)\d y,
\end{equation*}
where $c|U_f'(0)|=\mathcal{H}(0)$, due to Eq.~(\ref{frontsol}). Naturally, we recover that $\mathcal{E}(0)=0$ from the translation invariance of Eq.~(\ref{nfield}). Furthermore, $\mathcal{E}'(0)=\frac{1}{c\mathcal{H}(0)}\int_0^{+\infty}ye^{\frac{-y}{c}}w(y)\d y>0$, so $\lambda=0$ is a simple eigenvalue. In fact, it can be shown that $\lambda = 0$ is the only solution to $\mathcal{E}(\lambda) =0$.
Combining this with the fact that the essential spectrum always has negative real part, we conclude the traveling front solution $U_f$ is marginally stable~\cite{coombes04}. For the exponential kernel, Eq.~(\ref{exp}), we find $\mathcal{E}(\lambda)=\lambda/(c+1+\lambda)$~\cite{coombes04}, from which we clearly recover that $\lambda=0$ is the only solution of $\mathcal{E}(\lambda)=0$.

This concludes our analysis of entire solutions to Eq.~(\ref{nfield}) in the case $I \equiv 0$. Guided by the fact that the homogeneous solutions $\bar{u} \equiv 0,1$ are stable, and the intermediate bump $U_b(x)$ and periodic solutions $U_L(x)$ are unstable, we generally expect initial conditions $u_0(x)$ to either be attracted to $\bar{u}\equiv0$ or $\bar{u}\equiv1$ in the long time limit. In the next section, we demonstrate a means of determining the fate of unimodal initial conditions using interface equations.

\section{Nonequilibrium dynamics of a single active region}
\label{single}
In this section, we identify conditions on $u_0(x)$ with a single active region ($u_0(x) \geq \kappa$ for $x \in [x_1,x_2]$), so the solution to Eq.~(\ref{nfield}) propagates (assuming $I(x,t) \equiv 0$). In what follows, we assume $0\leq u_0(x) \leq 1$ is unimodal, $u_0'(x_0) = 0$ and $u_0'(x) \gtrless 0$ for $x \lessgtr x_0$, ensuring there are no more than two interfaces for $t>0$. First, we derive results for even $u_0(x) = u_0(-x)$, but our results can be extended to the case of asymmetric $u_0(x)$. Our analysis tracks the dynamics of Eq.~(\ref{nfield}) at the interfaces where $u(x_j(t),t) = \kappa$. Initial conditions can be separated into subthreshold ones that lead to decay and superthreshold ones that lead to propagation. Subsequently, we calculate asymptotic formulas from the interface equations for the extinction time of subthreshold solutions and the evolving propagation speed of superthreshold solutions. Lastly, we identify conditions on the external input $I(x,t)$ to Eq.~(\ref{nfield}) that ensure propagation when $u_0(x) \equiv 0$.

\subsection{Interface equations and criticality: even symmetric case}
\label{ifacecrit}
We start with smooth unimodal even initial conditions, $u_0(x) = u_0(-x)$, with a single active region, $u_0(x) \geq \kappa$ for $|x| \leq \ell$ and $u_0(x)< \kappa$ elsewhere, for $\ell > 0$ which satisfies $u_0'(x) \gtrless0$ for $x \lessgtr 0$. Symmetry of Eq.~(\ref{nfield}) with $I \equiv 0$ ensures solutions with even initial conditions are always even, so the active region $A(t) = \{ x\in\R~ | ~u(x,t) \geq \kappa \}$ will remain symmetric for $t> 0$.
The dynamics of the symmetric active region $A(t) = [-a(t), a(t)]$ can be described with interface equations for the two points $x= \pm a(t)$ (See \cite{amari77,coombes12}).
We start by rewriting Eq.~(\ref{nfield}) as
\begin{align}
\pd_t u(x,t) &= - u(x,t) + \int_{A(t)} w(x-y) \d y,
\label{afield}
\end{align}
which can be further simplified:
\[
\pd_t u(x,t) = -u(x,t)+W(x+a(t))-W(x-a(t)).
\]
Eq.~(\ref{afield}) remains well defined even in the case where $a(t)$ vanishes. We can describe the dynamics of the two interfaces by the implicit equations
\begin{align}
u(\pm a(t),t) = \kappa.  \label{dthresh}
\end{align}
Differentiating Eq.~(\ref{dthresh}) with respect to $t$, we find the total derivative is:
\begin{align}
\pm \alpha (t) a'(t) + \pd_t u(\pm a(t),t) = 0, \label{iface0}
\end{align}
where we define $a'(t)=\frac{\d a(t)}{\d t}$ and $\pm \alpha (t) = \pd_x u(\pm a(t),t)$. The symmetry of Eq.~\eqref{iface0} allows us to reduce to a single differential equation for the dynamics of $a(t)$:
\begin{align}
a'(t)= - \frac{1}{\alpha(t)} \left[ W(2a(t)) - \kappa \right],  \label{iface1}
\end{align}
where we have substituted Eq.~(\ref{afield}) at $a(t)$ for $\pd_t u(a(t),t)$. Eq.~\eqref{iface1} is not well-defined for $\alpha(t)=0$, but we will show how to circumvent this difficulty.  Furthermore, we can obtain a formula for $\alpha(t)$ by defining $z(x,t) : = \pd_x u(x,t)$ and differentiating Eq.~(\ref{afield}) with respect to $x$ to find~\cite{coombes12}
\begin{align*}
\pd_t z(x,t) &= - z(x,t) + w(x+a(t)) - w(x-a(t)), 
\end{align*}
which we can integrate and evaluate at $a(t)$ to find
\begin{align}
\alpha(t) = u_0'(a(t)) \e^{-t} + \e^{-t} \int_{0}^{t} \e^{s} \left[ w(a(t)+a(s)) - w(a(t)-a(s)) \right]  \d s.  \label{alpha1}
\end{align}
Thus, we have a closed system describing the evolution of the right interface $a(t)$ of the active region $A(t)$, given by Eqs.~(\ref{iface1}) and (\ref{alpha1}), along with the initial conditions $a(0) = \ell$ and $\alpha(0) = u_0'(\ell) < 0$, as long as $\alpha (t) < 0$. 
Criticality occurs for initial conditions such that $a'(t) = 0$, which means $W(2\ell)=\kappa$, {\it i.e.} for $\ell = b = W^{-1}(\kappa)/2$, so the critical $\ell$ is precisely the half-width of the unstable stationary bump solution $U_b(x)$ defined in Eq.~(\ref{statbump}). \\
\vspace{-3mm}

\begin{figure}
\begin{center} \includegraphics[width=13cm]{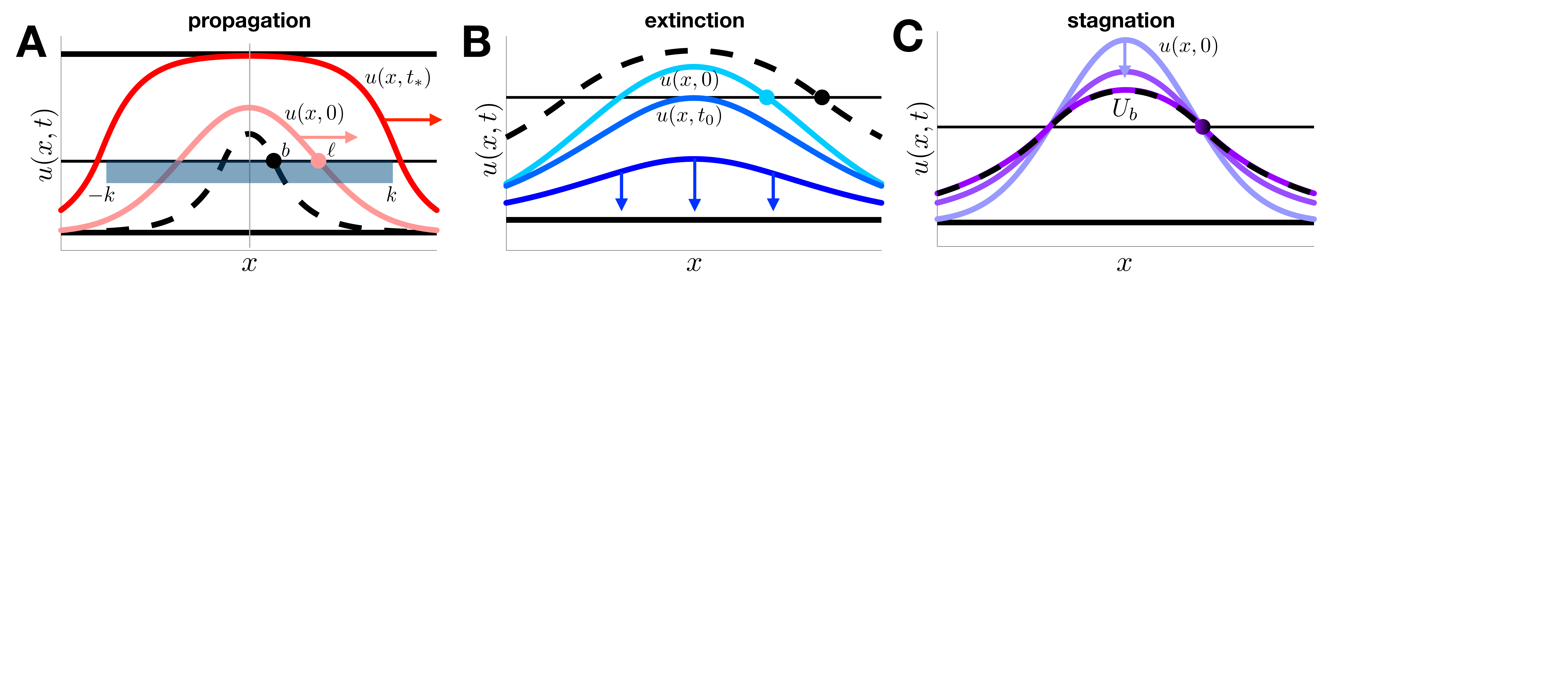} \end{center}
\vspace{-3mm}
\caption{Long term behavior of $u(x,t)$ depends only on how the initial interface location $a(0) = \ell$ compares to the bump half-width, $b = W^{-1}(\kappa)/2$. (A) If $\ell > b$, propagation occurs and $\lim_{t \to \infty}u(x,t) \equiv 1$, $\forall x \in K = [-k,k]$ for $k < \infty$. This follows from the fact that for any $K$, we can find a time $t_*$ for which $u(x,t_*)> \kappa$, $\forall x \in K$. (B) If $\ell < b$, eventually $u(x,t)< \kappa$, right after the time $t_0$ when $u(0,t_0) = \kappa$, and so $\lim_{t \to \infty} u(x,t) \equiv 0$. (C) If $\ell = b$, stagnation occurs and $\lim_{t \to \infty} u(x,t) = U_b(x)$. }
\vspace{-4mm}
\label{fig3_prop}
\end{figure}

\noindent
{\bf Propagation.} If $\ell>W^{-1}(\kappa)/2$ then $a'(t)>0$ and, due to the monotonicity of $w$ and Eq.~\eqref{alpha1}, $\alpha(t)<0$ for all time $t>0$ so $\lim_{t \to \infty} a(t) = \infty$, and the active region $A(t)$ expands indefinitely. As a consequence, for any compact set $K=[-k,k]$ with $k>0$ given and any $\epsilon>0$, we can find $t_*>0$ large enough such that $K\subset A(t_*)$ and 
\begin{align*}
\left| W(x+a(t_*))-W(x-a(t_*))-1 \right| \leq \epsilon, \quad \forall x\in K,
\end{align*}
so that for any equal or later time $s \geq t_*$ we have
\begin{align*}
\left| W(x+a(s))-W(x-a(s))-1 \right| \leq \epsilon, \quad \forall x\in K.
\end{align*}
We can solve for $u(x,t)$ starting for time $t_*$ to obtain
\begin{align*}
u(x,t)=u(x,t_*)e^{t_*-t}+e^{-t}\int_{t_*}^t e^s \left( W(x+a(s))-W(x-a(s)) \right)\d s.
\end{align*}
Using the fact that any solution is continuous, we have that $|u(x,t_*)|\leq M$ for all $x\in K$. As a consequence, we get that $\forall x\in K$,
\begin{align*}
|u(x,t)-1| &= \left| (u(x,t_*)-1)e^{t_*-t} + e^{-t}\int_{t_*}^t e^s \left( W(x+a(s))-W(x-a(s)) -1\right)\d s\right|\\
&\leq (1+M) e^{t_*-t} + \epsilon. 
\end{align*}
This implies that $\lim_{t \to \infty} |u(x,t)-1|= 0$, $\forall x\in K$. As a consequence, the solutions of Eq.~\eqref{nfield} locally uniformly converge to the homogeneous state $u\equiv1$ as $t \to \infty$ (Fig. \ref{fig3_prop}A). Thus, we have propagation of $u\equiv1$ into $u\equiv0$ as time evolves. \\
\vspace{-3mm}

\noindent
{\bf Extinction.} If $\ell<W^{-1}(\kappa)/2$, then $a'(t)<0$ and $0<a(t)<\ell$ on $t \in (0,t_0)$. By continuity, there exists a finite $t_0>0$ such that $a(t_0)=0$, at which point the interface dynamics, Eq.~(\ref{iface1}) and (\ref{alpha1}), breaks down. We know this because $W(2a(t)) - \kappa <0$ and decreases as $a(t)$ decreases. Note also that for $t\in(0,t_0)$ we consistently have $\alpha(t)<0$. Inspecting Eq.~\eqref{alpha1} shows that $\lim_{t \to t_0^-} \alpha(t) = 0$ since $u'_0(0)=0$. Thus, at time $t=t_0$, we have $0\leq u(x,t_0) \leq \kappa$, and for $t\geq t_0$, $\pd_t u(x,t) = -u(x,t)$, so $u(x,t)=e^{t_0-t}u(x,t_0)$ for $t\geq t_0$, and $\lim_{t \to \infty}u(x,t)=0$, uniformly on $x \in \R$ (Fig. \ref{fig3_prop}B). \\
\vspace{-3mm}

\noindent
{\bf Stagnation.} If $\ell= W^{-1}(\kappa)/2$, then $a'(t)=0$ for all time assuming $\alpha(t)<0$, implying $a(t) \equiv \ell$. Plugging into Eq.~\eqref{alpha1} yields $\alpha(t)= (w(2b)-w(0))(1-\e^{-t})+ u_0'(\ell)\e^{-t} <0$ for $t>0$. As a consequence, $a(t)=\ell$ for all time and $\lim_{t \to \infty} \alpha(t) =  w(2b) - w(0)$. Furthermore, we can explicitly solve for
\begin{align*}
u(x,t)& = W(x+b)-W(x-b)+e^{-t}\left[ u_0(x) - W(x+b)+W(x-b)\right],
\end{align*}
so $\lim_{t \to \infty} u(x,t) = U_b(x)$, uniformly on $\R$. We call this case stagnation as the active region remains fixed for $t>0$ (Fig. \ref{fig3_prop}C).

To summarize, we have shown the following result. \\
\vspace{-4mm}

{\em Starting with smooth unimodal even initial conditions, $u_0(x) = u_0(-x)$, with a single active region, $u_0(x) \geq \kappa$ for $|x| \leq \ell$ and $u_0(x)< \kappa$ elsewhere, $\ell > 0$ satisfying $u_0'(x) \gtrless0$ for $x \lessgtr 0$, the fate of the solutions $u(x,t)$ to the Cauchy problem, Eq~\eqref{nfield}, falls into three cases: \\
\vspace{-4mm}

\begin{itemize}
\item[(i)]  If $\ell>W^{-1}(\kappa)/2$, then $u\rightarrow 1$ locally uniformly on $\R$ as $t\rightarrow +\infty$;
\item[(ii)]  If $\ell<W^{-1}(\kappa)/2$, then $u\rightarrow 0$  uniformly on $\R$ as $t\rightarrow +\infty$;
\item[(iii)]  If $\ell= W^{-1}(\kappa)/2$, then $u\rightarrow U_b$  uniformly on $\R$ as $t\rightarrow +\infty$.
\end{itemize}}

\subsection{Interface equations and criticality: asymmetric case} We can extend our analysis to the case of initial conditions $u_0(x)$ that are still unimodal, $u_0'(x_0) = 0$ with $u_0'(x) \gtrless 0$ for $x \lessgtr 0$, but can be asymmetric, so $u_0(x) \neq u_0(-x)$ for some $x$. Conditions can be stated in terms of the active region of the initial condition $A(0) = \left[ \bar{x}_1, \bar{x}_2 \right]$, where $u_0(x) \geq \kappa$. The active region of $u(x,t)$ is now defined $A (t) = [x_1(t), x_2(t)]$ with associated spatial gradients $\alpha_j(t) = \pd_xu(x_j(t),t)$ for $j=1,2$. Carrying out a derivation of the interface dynamics then yields~\cite{coombes12,krishnan17}
\begin{subequations} \label{ifaceasym}
\begin{align}
x_j'(t) &= - \frac{1}{\alpha_j(t)} \left[ W(x_2(t) - x_1(t)) - \kappa \right], \\
\alpha_j(t) &= u_0'(x_j(t)) \e^{-t} + \e^{-t} \int_0^t \e^s \left[ w(x_j(t) - x_1(s)) - w(x_j(t) - x_2(s)) \right] \d s,
\end{align}
\end{subequations}
along with initial conditions $x_j(t) = \bar{x}_j$ and $\alpha_j(0) = u_0'(\bar{x}_j)$ for $j=1,2$, now requiring $\alpha_1(t)>0$ and $\alpha_2(t)< 0$. Criticality occurs for initial conditions such that $x_j'(t) = 0$, which means $W(\bar{x}_2 - \bar{x}_1) = \kappa$, so the critical width $2b : = W^{-1} ( \kappa)$ is precisely the width of the stationary bump $U_b(x)$. Similar to our findings in the symmetric case, we can show: (i) propagation occurs if $\bar{x}_2 - \bar{x}_1 > 2b $; (ii) extinction occurs for $\bar{x}_2 - \bar{x}_1 < 2b$; and (iii) stagnation occurs for $\bar{x}_2 - \bar{x}_1 =2b$.

\subsection{Asymptotic results}
\label{asymexp}
As demonstrated, we can predict the long term dynamics of Eq.~(\ref{nfield}) based on the initial condition $u_0(x)$ and a subsequent analysis of the interface dynamics. The interface equations also allow for the derivation of some convenient asymptotic approximations. In particular, we can estimate speed of propagating solutions in the long time limit, showing they are consistent with our results for traveling fronts. We also estimate the time to extinction of decaying solutions. To do so, we carry out a truncation to leading order of the system of Eq.~(\ref{iface1}) and (\ref{alpha1}) in the symmetric case. \\
\vspace{-3mm}

\noindent
{\bf Long term propagation speed.} For propagating solutions, we know $\lim_{t \to \infty} a(t) = +\infty$. Assuming the interface propagates at constant speed $a(t) \sim ct + a_0$ in the limit $t \to \infty$, self-consistency is enforced by plugging into Eq.~(\ref{alpha1}) and evaluating
\begin{align*}
\lim_{t \to \infty} \alpha (t) &= \lim_{t \to \infty} \left[ u_0'(a(t)) \e^{-t} + \int_0^t \e^{-(t-s)} \left[ w(c(t+s)+2a_0) - w(c(t-s)) \right] \d s \right] \\
&= - \lim_{t \to \infty} \int_0^t \e^{-(t-s)} w(c(t-s)) \d s = - \frac{1}{c} \int_0^{\infty} \e^{-y/c} w(y) \d y : = \bar{\alpha}.
\end{align*}
Differentiating Eq.~(\ref{frontsol}) for $U_f(\xi)$ and plugging in $\xi = 0$, we obtain the same formula, so $\bar{\alpha} = U_f'(0)$, the gradient of the traveling front solution at the threshold $\kappa$. Plugging into Eq.~(\ref{iface1}) along with our assumption $a(t) = ct + a_0$, we find an implicit equation for $c$, $\int_0^{\infty} \e^{-y/c} w(y) \d y = W_{\infty} - \kappa$, which matches Eq.~(\ref{frontthresh2}). \\
\vspace{-3mm}

\noindent
{\bf Time to extinction.} To approximate the extinction time $t_0$ when $a(t_0) = 0$ in the case $\ell < W^{-1} (\kappa)/2$, we work in the limit $0<\ell \ll 1$. As $0<a(t)<\ell$ for time $t\in(0,t_0)$, a Taylor expansion of Eqs.~\eqref{iface1} and \eqref{alpha1} in $0< a(t) \ll 1$ implies $\alpha (t)$ and $t_0$ are small too.
In this case, we can approximate $\alpha (t) \approx u_0''(0) \ell$, using the leading order term in Eq.~(\ref{alpha1}), so plugging into Eq.~(\ref{iface1}) and integrating we can estimate the extinction time $t_0$ as:
\begin{align}
\frac{\ell}{\kappa} \approx \frac{t_0}{\ell |u_0''(0)|} \ \ \Rightarrow \ \ t_0 \approx \ell^2 |u_0''(0)|/\kappa \text{ as } \ell \rightarrow 0.  \label{exttime}
\end{align}
\ \\
\vspace{-5mm}

\subsection{Critical stimulus for activation}
\label{critstim}
We now consider the impact of spatiotemporal inputs $I(x,t)$ on the long term dynamics of Eq.~(\ref{nfield}) when $u_0(x) \equiv 0$. This may be more biologically realistic than assuming arbitrary initial conditions, as waves are often initiated experimentally in quiescent neural tissue via the application of a brief external stimulus~\cite{delaney94,ferezou06,xu07}. To provide intuition, we first construct stationary solutions assuming $I(x,t) \equiv I(x)$ is unimodal ($I'(0) = 0$ and $I'(x) \gtrless 0$ for $x \lessgtr 0$), positive $I(x) > 0$, and even $I(x) = I(-x)$. When $\max_{x \in \R} I(x) = I(0) > \kappa$, we show that if there are any stationary bump solutions, the one with minimal half-width $b_{\rm min}$ is linearly stable. Subsequently, we derive conditions for a brief stimulus lasting a time $t_1$, $I(x,t) = I(x)\chi_{[0,t_1]}$ ($\chi_{[0,t_1]} =1$, $t \in [0,t_1]$; 0 otherwise), that ensure propagation of solutions for times $t> t_1$. We show that: (i) there must be no stationary bump solutions to Eq.~(\ref{nfield}) with $I(x,t) = I(x)$; and (ii) the active region at $t=t_1$ must be wider than that of the critical bump $U_b(x)$ of the input-free system.

\begin{figure}
\begin{center} \includegraphics[width=13cm]{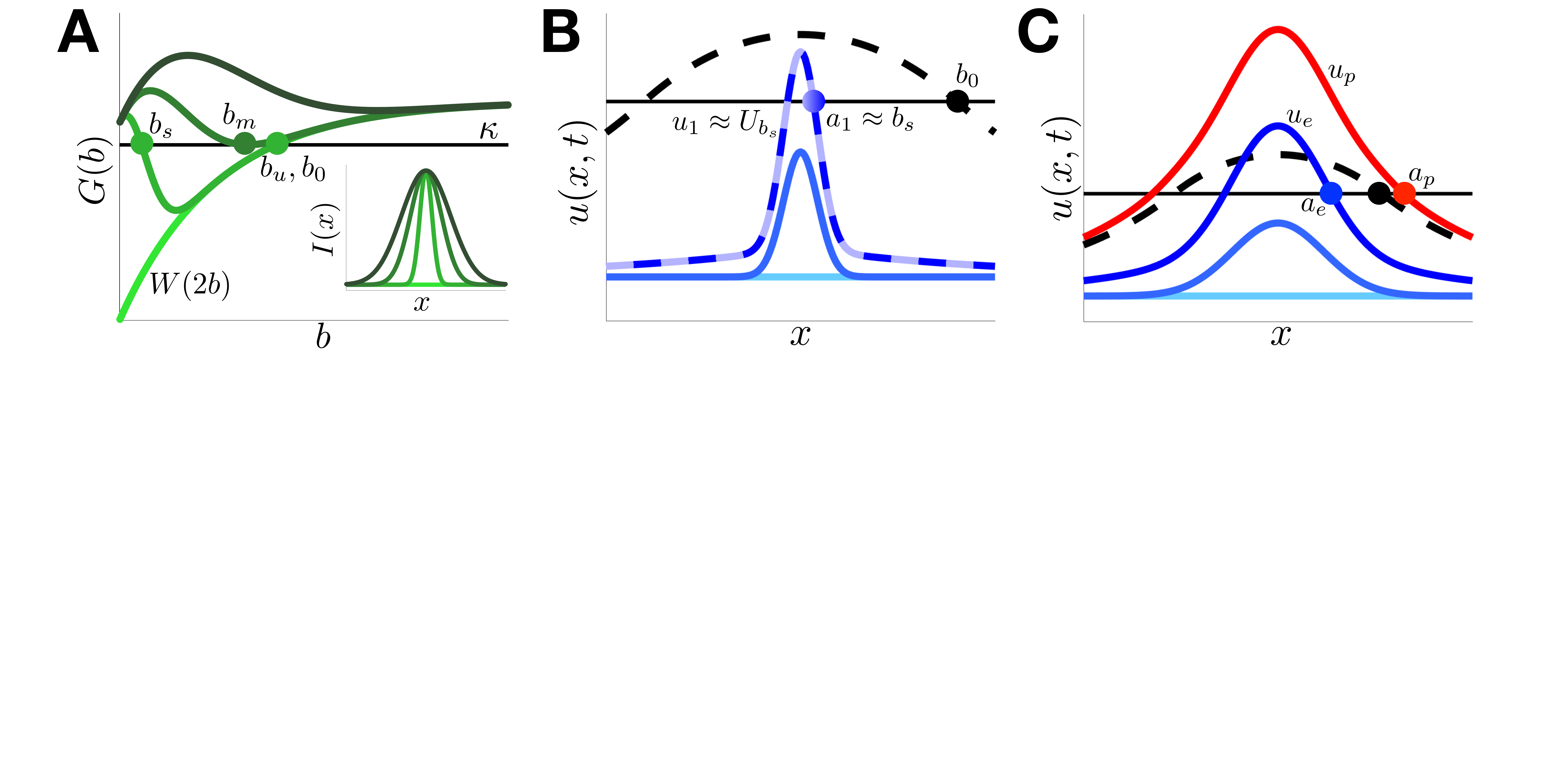} \end{center}
\vspace{-2mm}
\caption{Conditions for propagation driven by the input $I(x,t) = I(x)\chi_{[0,t_1]}$, when $u(x,0) \equiv 0$. (A)  For symmetric, unimodal, and positive profile $I(x)$ with $I(0) > \kappa$, propagation will only occur if $G(b) = W(2b)+I(b) = \kappa$ has no solutions, which occurs for sufficiently wide $I(x)$ (note inset). If solutions to Eq.~(\ref{inpbump}) exist, the minimal one will be linearly ($b_s$) or marginally ($b_m$) stable. Corresponding unstable solutions $b_u$ will typically be close to $b_0$, the solution to $W(2b_0) = \kappa$. (B) For $I(x)$ such that $G(b_s) = \kappa$ for some $b_s$, $u(x,t_1) \approx U_{b_s}(x)$ for large enough $t_1$ with active region $[-a_1,a_1]$ for $a_1 : = a(t_1) < b_0$, so $\lim_{t \to \infty} u(x,t) \equiv 0$. (C) Here, $I(x)$ is chosen such that $G(b) = \kappa$ has no solutions. Taking $t_e$ such that $u_e = u(x,t_e)$ satisfies $u(\pm a_e, t_e) = \kappa$ with $a_e < b_0$, then $\lim_{t \to \infty} u(x,t) \equiv 0$. On the other hand, for $t_p$ such that $u_p = u(x,t_p)$ satisfies $u(\pm a_p, t_p) = \kappa$ with $a_p > b_0$, then $u(x,t)$ propagates as $t \to \infty$.}
\vspace{-4mm}
\label{fig4_input}
\end{figure}

Stationary bump solutions to Eq.~(\ref{nfield}) for $I(x,t) \equiv I(x)$ with a single active region have the form $U_b(x) = W(x+b) - W(x-b) + I(x)$. The threshold condition
\begin{align}
U_b(\pm b) = W(2b) + I(b) = G(b) = \kappa \label{inpbump}
\end{align}
defines an implicit equation for the half-width $b$. If there are solutions $b$ to Eq.~(\ref{inpbump}), they will all be less than the solution to the input-free case $I \equiv 0$: $b<b_0 = W^{-1}(\kappa)/2$. To show this, we subtract Eq.~(\ref{inpbump}) from the equation $W(2b_0) = \kappa$ to find $W(2b_0) - W(2b) = I(b) > 0$, so $W(2b_0) > W(2b)$, and since $W(x)$ is monotone increasing then $b_0 > b$. See Fig. \ref{fig4_input}A for illustration. Local stability is characterized as before, by deriving a linearized equation for the perturbation $u(x,t) = U_b(x) + \e^{\lambda t} \psi (x)$. For odd perturbations $\psi (b) = - \psi (-b)$, the associated eigenvalue
\[
\lambda_o = I'(b)/(w(0) - w(2b) - I'(b)) <0,
\]
since $I(b)$ is decreasing for $b>0$. For even perturbations $\psi (b) = \psi (-b)$, we find
\begin{align*}
\lambda_e = \frac{2w(2b) + I'(b)}{w(0) - w(2b) - I'(b)} = \frac{G'(b)}{w(0) - w(2b) - I'(b)},
\end{align*}
so the sign of $\lambda_e$ is the same as the sign of $G'(b)$. The bump will thus be stable if $G'(b)<0$. When $I(0)> \kappa$, we know $G(0) > \kappa$. Since $G(x)$ is continuous, if there are any solutions to Eq.~(\ref{inpbump}), the minimal one $b_{\rm min}$ will be stable or marginally stable, since $G(x)$ will be decreasing or at a local minimum there. A similar analysis was performed for a neural field model with linear adaptation in \cite{folias04}, but the adaptation could also induce oscillatory instabilities.

We now demonstrate that for a spatiotemporal input, $I(x,t) = I(x) \chi_{[0,t_1]}$, to generate propagation, (i) Eq.~(\ref{inpbump}) must have no solutions, and (ii) $t_1$ must be large enough so the active region $A(t) = [-a(t), a(t)]$ satisfies $a(t_1)>b_0$, where $b_0$ solves Eq.~(\ref{inpbump}) for $I \equiv 0$. Starting from $u_0(x) \equiv 0$, we know initially, the dynamics obeys $\pd_t u(x,t) = -u(x,t) + I(x,t)$, so $u(x,t) = I(x) (1 - \e^{-t})$ during this phase. This formula determines the lower bound on the stimulus time $t_0<t_1$ needed to generate a nontrivial active region, $A(t) \neq \varnothing$. This time is given by solving 
\[
\max_{x \in \R} u(x,t_0) = I(0) (1 - \e^{-t_0}) = \kappa \ \ \ \Rightarrow \ \ \ t_0 = \ln \left[ \frac{I(0)}{I(0) - \kappa} \right].
\]
If $t_1\leq t_0$, then the long term dynamics of the solution is $u(x,t) = I(x)(1 - \e^{-t_1}) \e^{-(t-t_1)}$ for $t> t_1$, and $\lim_{t \to \infty} u(x,t) \equiv 0$. Note if $I(0) < \kappa$, then $u(x,t) < \kappa$ for all $t>0$.

If $t_1 > t_0$, then for $t_0< t < t_1$, we can derive the interface equations for $u(\pm a(t), t) = \kappa$, which are
\begin{subequations} \label{singinpface}
\begin{align}
a'(t) &= - \frac{1}{\alpha(t)} \left[ W(2a(t)) - \kappa + I(a(t)) \right], \\
\alpha (t) &= \e^{-t} \int_{t_0}^{t} \e^s \left[ w(a(t) + a(s)) - w(a(t) - a(s)) + I'(a(s)) \right] \d s,
\end{align}
\end{subequations}
with initial conditions $a(t_0) = 0$ and $\alpha(t_0) = 0$, so $a'(t_0)$ diverges. Despite the singularity, we can show that $a'(t)$ is integrable for $|t-t_0| \ll 1$ and $a(t), \alpha (t) \propto \sqrt{t - t_0}$. We desingularize Eq.~(\ref{singinpface}) with the change of variables $\tau = - \int_{t_0}^t \frac{\d s}{\alpha (s)}$~\cite{aronson80,avitabile17}, so the differential equation for $\tilde{a}(\tau)$ in the new coordinate frame is
\begin{align}
\frac{\d \tilde{a}}{\d \tau}(\tau) = W(2 \tilde{a}(\tau)) - \kappa + I(\tilde{a}(\tau)),  \label{dsinpface}
\end{align}
with $\tilde{a}(0) = 0$. Since we know $\alpha (t) < 0$ for $t> t_0$, then $\tau$ will be an increasing function of $t$, so we refer now to $\tau_1 : = \tau (t_1)$ and note $0 = \tau (t_0)$. Because $I(0) - \kappa>0$ by assumption, we have $\frac{\d \tilde{a}}{\d \tau}(\tau) > 0$ for all $\tau$ where it is defined.

There are three remaining cases now, which depend on the existence of solutions to Eq.~(\ref{inpbump}) and the time $\tau_1 >0$: (I) Eq.~(\ref{inpbump}) has at least one solution, and propagation does not occur; (II) Eq.~(\ref{inpbump}) has no solutions, but $\tau_1 \leq \tau_c$, the time at which $\tilde{a}(\tau_c) = b_0$ for $I(x, \tau) \equiv I(x)$, and propagation does not occur; (III) Eq.~(\ref{inpbump}) has no solutions, and $\tau_1 > \tau_c$, so propagation occurs. We now treat these three cases in detail. \\
\vspace{-3mm}

\noindent
{\em Case I: $\min_{x \in \R} G(x) \leq \kappa$.} Here, Eq.~(\ref{inpbump}) possesses at least one solution. By our assumption $I(0)> \kappa$, this solution $b_{\rm min}$ is linearly or marginally stable, as we have shown. Eq.~(\ref{dsinpface}) implies $\frac{\d \tilde{a}}{\d \tau} > 0$ for all $\tau < \tau_1$, but $\frac{\d \tilde{a}}{\d \tau}$ vanishes at $\tilde{a} = b_{\rm min}$, so $\tilde{a}(\tau) < b_{\rm min} < b_0$ for all $\tau < \tau_1$. Thus, once $\tau = \tau_1$, the dynamics is described by the extinction case detailed in Section \ref{ifacecrit}, and $\lim_{t \to \infty} u(x,t) \equiv 0$ (Fig. \ref{fig4_input}B). \\
\vspace{-3mm}

\noindent
{\em Case II: $\min_{x \in \R} G(x)>\kappa$ and $\tau_1 \leq \tau_c$.} Here Eq.~(\ref{inpbump}) has no solutions, but $\tilde{a}(\tau)$ will not grow large enough for propagation to occur once the input $I(x,\tau)$ is terminated. This is due to the condition $\tau_1 \leq \tau_c$, where we can define the critical time $\tau_c$ as the time when $\tilde{a}(\tau_c) = b_0 = W^{-1}(\kappa)/2$ as
\begin{align}
\int_0^{W^{-1}(\kappa)/2} \frac{\d a}{W(2a) - \kappa + I(a)} = - \int_{t_0}^{t_c} \frac{\d t}{\alpha (t)} : = \tau_c. \label{tauc}
\end{align}
By definition $\tilde{a}(\tau_1) \leq b_0$, so once $\tau = \tau_1$, the dynamics is described by (a) the extinction case in Section \ref{ifacecrit} if $\tau_1 < \tau_c$, so $\lim_{t \to \infty} u(x,t) \equiv 0$, or (b) the stagnation case in Section \ref{ifacecrit} if $\tau_1 = \tau_c$, so $\lim_{t \to \infty} u(x,t) \equiv U_b(x)$ (Fig. \ref{fig4_input}C).  \\
\vspace{-3mm}

\noindent
{\em Case III: $\min_{x \in \R} G(x)>\kappa$ and $\tau_1 > \tau_c$.} Finally, we describe the case ensuring propagation for $t \to \infty$. Requiring $\tau_1> \tau_c$ with Eq.~(\ref{tauc}), we have that $\tilde{a}(\tau_1)> b_0$. After $\tau = \tau_1$, the dynamics is described by the propagation case in Section \ref{ifacecrit}, so the homogeneous state $u \equiv 1$ is locally uniformly propagating as $t \to \infty$ (Fig. \ref{fig4_input}C).

\begin{figure}
\begin{center} \includegraphics[width=10cm]{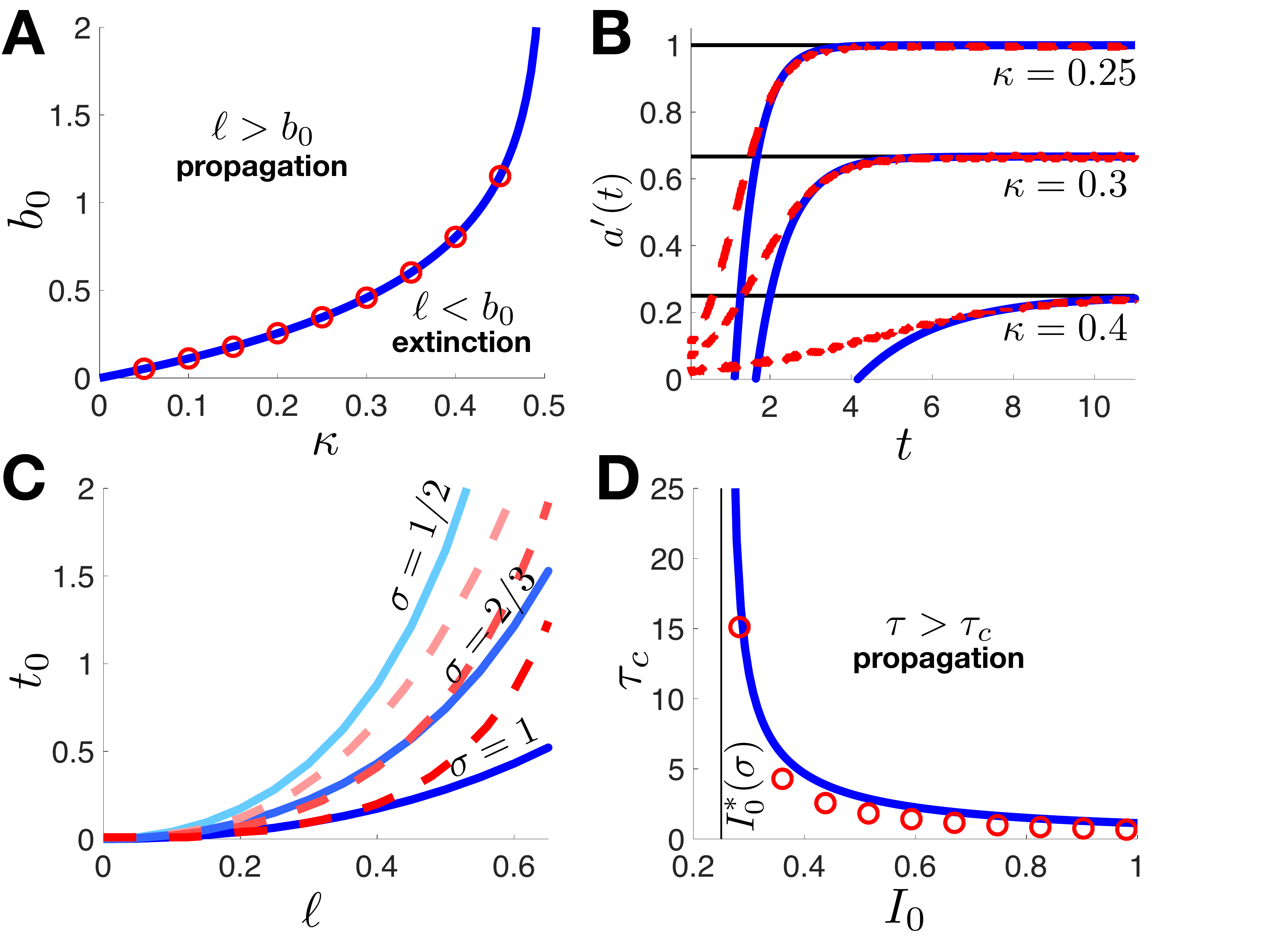} \end{center}
\vspace{-3mm}
\caption{Bounds and asymptotics for an exponential kernel, $w(x) = \e^{-|x|}/2$. (A) Propagation occurs if $\ell : = a(0) > b_0$ for $b_0 = - \ln \left[1-2 \kappa \right]/2$ (solid line); extinction occurs if $\ell < b_0$. Numerical simulations (circles) determine $b_0$ by computing Eq.~(\ref{nfield}) starting with $u(x,0) = u_0(x)$ and identifying $b_0$ for which stagnation occurs ($u_0(b_0) = \kappa$). Results are consistent whether choosing $u_0(x) = {\mc U} \cdot U_b(x), {\mc U}\e^{-x^2},$ or $ \left[ {\mc U} (1-x^2) \right]_+$. (B) Instantaneous speed of interface $a'(t) \to c(\kappa)$ in numerical simulations (dashed lines). For $a'(t) \approx c$, the approach is well characterized by the asymptotic estimate $a'(t) \approx c - c_1 \e^{-2 ct}$ (solid line) for best fit $c_1$. (C) Extinction time $t_0 \approx \ell \e^{\ell^2/(2 \sigma^2)}/\sigma^2$ (solid line) estimated for $u_0(x) = {\mc U} \e^{-x^2/(2\sigma^2)}$ compared with numerical simulations. (D) Critical time $\tau_c$ (in rescaled coordinate $\tau = -\int_{t_0}^{t} \frac{\d s}{\alpha(s)}$) the input $\tilde{I}(x,\tau) = I_0 \chi_{[0,\tau_1]} \e^{-5|x|}$ must be on for propagation to occur, computed from Eq.~(\ref{tauc}) by integrating in $a$ using quadrature (solid line) or computing Eq.~(\ref{nfield}) and numerically computing the integral in $t$ (circles). As $I_0 \to I_0^*(\sigma)$, the minimal $I_0$ for propagation, $\tau_c$ blows up.}
\vspace{-5mm}
\label{fig5_expwt}
\end{figure}

\subsection{Explicit results for exponential kernel} Lastly, we demonstrate the results derived above by employing the judicious chosen exponential kernel, Eq.~(\ref{exp}). The form of the interface equations for symmetric initial conditions and $I \equiv 0$ are
\begin{subequations} \label{expiface}
\begin{align}
a'(t) &= - \frac{1}{2 \alpha (t)} \left[ 1 - \e^{-2a(t)} - 2\kappa \right], \\
\alpha(t) &= u_0'(a(t)) \e^{-t} - \e^{-t - a(t)} \int_0^t \e^s \sinh(a(s)) \d s.
\end{align}
\end{subequations}
First, note the critical half-width $b_0$ is given by when $a'(t) = 0$, which here is $b_0 = - \frac{1}{2} \ln \left[1-2 \kappa \right] $, so if $a(0)>b_0$, propagation occurs. We demonstrate the accuracy of this boundary in predicting long-term dynamics by comparing with numerically computed boundaries in Fig. \ref{fig5_expwt}A. Note, in the case of propagation, in the limit $t \gg 1$, we can approximate $a(t) \approx ct + a_0$, and the asymptotic approximation in Section \ref{asymexp} yields $\frac{c}{2(c+1)} = \frac{1}{2} - \kappa$, which we rearrange to yield~\cite{ermentrout93,pinto01,bressloff01}
\[
c = \frac{1}{2 \kappa} \left[ 1 - 2 \kappa \right], \ \ \ \ \ \bar{\alpha} = - \frac{2 \kappa}{1 - 2 \kappa} \cdot \frac{1 - 2 \kappa}{2} = - \kappa.
\]
To quantify the timescale of approach to the asymptotic dynamics, we study the evolution of perturbations to the long term wavespeed $c$, $a(t) = ct + a_0 + \phi (t)$ and assuming $\alpha (t) \approx - \kappa$. Plugging into Eq.~(\ref{expiface}), and truncating assuming $\phi (t)$ and $\e^{-2a_0}$ are of similar order, we find
\[
2 \kappa \phi'(t) = \e^{-2(ct+a_0)} \ \ \ \Rightarrow \ \ \ \phi (t) = - \frac{\e^{-2 a_0}}{ 4 \kappa c} \e^{-2ct},
\]
and $a(t)$ approaches the propagation speed $c$ at rate $2c$. We compare this result to our findings from numerical simulations in Fig. \ref{fig5_expwt}B. We save a higher order asymptotic analysis for future work. In addition, we can compute the asymptotic extinction time for the case in which $u_0(x) = {\mc U} \e^{-x^2/(2\sigma^2)}$, so $|u_0''(0)| = {\mc U}/ \sigma^2$ and $\kappa = u_0(\ell)$ implies
\[
\ell = \sigma \sqrt{2} \sqrt{\ln ({\mc U}/ \kappa)} \ \ \ \Rightarrow \ \ \ t_0 \approx \ell^2 \e^{\ell^2/(2\sigma^2)}/\sigma^2,
\]
which agrees with numerical simulations for small enough $\ell$ (Fig. \ref{fig5_expwt}C).

The critical stimulus for activation was determined in for a general weight kernel in Section \ref{critstim}. Note, the main conditions are that Eq.~(\ref{inpbump}) has no solutions, and that the stimulus remains on for a time $t>t_c$, where $t_c$ is defined by the relation in Eq.~(\ref{tauc}). We can gain more insight from these equations by studying the case of an exponential weight kernel, Eq~(\ref{exp}), and an exponential input $I(x) = I_0 \e^{-|x|/\sigma}$ (See \cite{folias04} for analysis of a related model with $I(x) = I_0 \e^{-x^2/(2 \sigma^2)}$). Thus, Eq.~(\ref{inpbump}) becomes
\[
\kappa = (1 - \e^{-2b})/2 + I_0 \e^{-b/ \sigma} = G(b),
\]
so
\[
G'(b) = \e^{-2b} - \frac{I_0}{\sigma} \e^{-b/ \sigma} = 0 \ \ \ \Rightarrow \ \ \ b^* = \sigma \ln \left[ I_0/ \sigma \right]/(1 -2 \sigma)
\]
and also $\lim_{b \to 0^+} G'(b) = 1- I_0/ \sigma$. Therefore if the input is sufficiently wide, $\kappa < I_0 <  \sigma$ and $1/2 < \sigma$, then initially $G(b)$ increases until $b^* > 0$, and then it decreases to $1/2$ for large $b$, so $G(b)> \kappa$ for all $b>0$ for sufficiently wide inputs with $I_0 > \kappa$. In addition, even for $I_0 > \sigma > 1/2$, then $b^*< 0$, and since we know $\lim_{b \to \infty} G(b) = 1/2$, then $G(b) > 1/2$ since it must be monotone decreasing for all $b>0$. Thus, there are no stable bump solutions to Eq.~(\ref{inpbump}) for sufficiently wide and strong inputs. On the other hand, if we wish to determine the critical curve $I_0^* (\sigma)$ below which bump solutions to Eq.~(\ref{inpbump}) emerge (assuming $I_0 > \kappa$), we simultaneously solve $G(b)=\kappa$ and $G'(b) = 0$ to find the saddle-node bifurcation point
\[
I_0^* (\sigma) = \sigma \frac{1-2 \sigma}{1 - 2 \kappa} \e^{(1- 2 \sigma)/(2\sigma)}.
\]
Taking $I_0 \leq I_0^*(\sigma)$ then ensures the existence of bumps (as in Fig. \ref{fig4_input}A). For $I_0 > I_0^*(\sigma)$, we can also study the impact of the input on the time necessary to reach $a(t) = b_0$, using the integral over $a$ in Eq.~(\ref{tauc}). We evaluate this numerically in Fig. \ref{fig5_expwt}D, showing it compares well with estimates we obtain by computing the critical time $t_c$ numerically and then converting to $\tau$ coordinates using the change of variables in Eq.~(\ref{tauc}). Note that as $I_0 \to I_0^*(\sigma)$, then $\tau_c \to \infty$.

\section{Multiple active regions}
\label{multiple}
Having determined the critical choices of unimodal initial conditions $u(x,0) = u_0(x)$ and inputs $I(x,t)$ that ensure propagation, we now turn our attention to a more general case of multimodal initial conditions. Since this can now lead to multiple disjoint active regions (where $u_0(x) \geq \kappa$), we must extend our analysis from Section \ref{single} to track more than two interfaces (See also \cite{krishnan17}). While it is difficult to analyze the resulting system of equations explicitly, we can gain insight by focusing on two specific cases of $u_0(x)$: (a) periodic initial conditions having an infinite number of active regions and (b) two symmetric active regions. We begin by deriving the interface equations in the general case.

\subsection{Interface equations: general case}
We now extend our analysis to the case where the relation $u_0(x) \geq \kappa$ is satisfied by multiple disjoint active regions: $A(0) = \cup_{j=1}^N \left[ a_j(0), b_j(0) \right]$, so the time evolution of $A(t)$ is implicitly described by
\begin{align}
u(a_j(t),t) = u(b_j(t),t) = \kappa, \ \ \ \ j=1,...,N,  \label{multiface}
\end{align}
for an initial time $0<t<t_0$. Differentiating Eq.~(\ref{multiface}) with respect to $t$, we find
\begin{align}
\alpha_j(t) a_j'(t) + \pd_t u(a_j(t),t) = 0, \hspace{3mm} \beta_j(t) b_j'(t) + \pd_t u(b_j(t),t) = 0, \hspace{4mm} j=1,...,N, \label{multitder}
\end{align}
where $\alpha_j (t) = \pd_x u(a_j(t),t)$ and $\beta_j(t) = \pd_x u(b_j(t),t)$. Rearranging Eq.~(\ref{multitder}), applying Eq.~(\ref{afield}) for $u_t$, and solving for $z= u_x$ as before, we find the following system describing the evolution of the interfaces $(a_j(t),b_j(t))$ and gradients $(\alpha_j(t),\beta_j(t))$:
\begin{subequations} \label{multifacesys}
\begin{align}
a_j'(t) &= - \frac{1}{\alpha_j (t)} \left[ \sum_{k=1}^{N} \left( W(b_k(t) -a_j(t)) - W(a_k(t) - a_j(t)) \right) - \kappa \right], \\
b_j'(t) &= - \frac{1}{\beta_j(t)} \left[ \sum_{k=1}^N \left( W(b_k(t) - b_j(t)) - W(a_k(t) - b_j(t)) \right) - \kappa \right], \\
\alpha_j(t) &= \e^{-t} \int_0^t \e^s \sum_{k=1}^N \left[ w(a_j(t) - a_k(s))-w(a_j(t)-b_k(s))) \right] \d s + u_0'(a_j(t)) \e^{-t}, \\
\beta_j(t) &= \e^{-t} \int_0^t \e^s \sum_{k=1}^N \left[ w(b_j(t) - a_k(s)) - w(b_j(t) - b_k(s)) \right] \d s + u_0'(b_j(t)) \e^{-t},
\end{align}
\end{subequations}
for $j=1,...,N$. The initial conditions $u_0(a_j(0)) = u_0(b_j(0)) = \kappa$ close the system. We expect $\alpha_j(t) \geq 0$ and $\beta_j(t) \leq 0$, since they are at the left and right boundaries of each active region. For the system Eq.~(\ref{multifacesys}), there is no straightforward condition that will ensure propagation in all cases (e.g., see Fig.~\ref{fig1_separatrix}C). For $N=1$, Eq.~(\ref{multifacesys}) reduces to Eq.~(\ref{ifaceasym}), and recall we can explicitly compute the condition for propagation.

Despite the difficulty in generalizing our approach to ensuring propagation in the single active region case to multiple active regions, as described by Eq.~(\ref{multifacesys}), we can make analytical progress in some special cases. Furthermore, one can solve Eq.~(\ref{multifacesys}) much faster numerically than Eq.~(\ref{nfield}), allowing a computational route to identifying conditions on $u_0(x)$ that determine propagation. We save such numerical computations for future work. Here, we will focus on two special choices of initial conditions that admit further explicit analysis: initial conditions that are (a) periodic and (b) even symmetric with two active regions.

\subsection{Periodic initial conditions}
\label{peric}
We can leverage results on periodic stationary solutions derived in Section \ref{persolns} along with the analysis for single active regions in Section \ref{ifacecrit} to derive conditions for saturation ($u \to 1$) when initial conditions are periodic. For an even and periodic initial condition $u(x,0) = u_L(x)$ of period $L$, $A(t) = \cup_{n \in \Z} \left[ -a(t) + nL, a(t) + nL \right]$, so by symmetry we can reduce Eq.~(\ref{multifacesys}) to
\begin{subequations} \label{periface}
\begin{align}
a'(t) &= - \frac{1}{\alpha(t)} \left[ {\bf W}_L(a(t)) - \kappa \right], \\
\alpha (t) &= u_L'(a(t)) \e^{-t} + \e^{-t} \int_0^t \e^{s} \sum_{n \in \Z} w_n(a(t),a(s)) \d s,
\end{align}
\end{subequations}
where $w_n(a(t),a(s)) = w(a(t)+a(s) + nL) - w(a(t)-a(s) + nL)$ and ${\bf W}_L(x)$ is defined as in Eq.~(\ref{WL}). Fixing $L$, the initial condition $u_L(x)$ is defined by the single parameter $\ell_L : = a(0)$, where $u_L(\pm \ell_L + nL) = \kappa$, $\forall n \in \Z$. Criticality occurs for $\ell_L = b_L(\kappa) = {\bf W}_L^{-1}(\kappa)$, the half-width of each active region of the periodic solution $U_L(x)$, Eq.~(\ref{statper}). The analysis proceeds along similar lines to that given in Section \ref{ifacecrit} for the single active region case. \\
\vspace{-3mm}

\begin{figure}
\begin{center} \includegraphics[width=13cm]{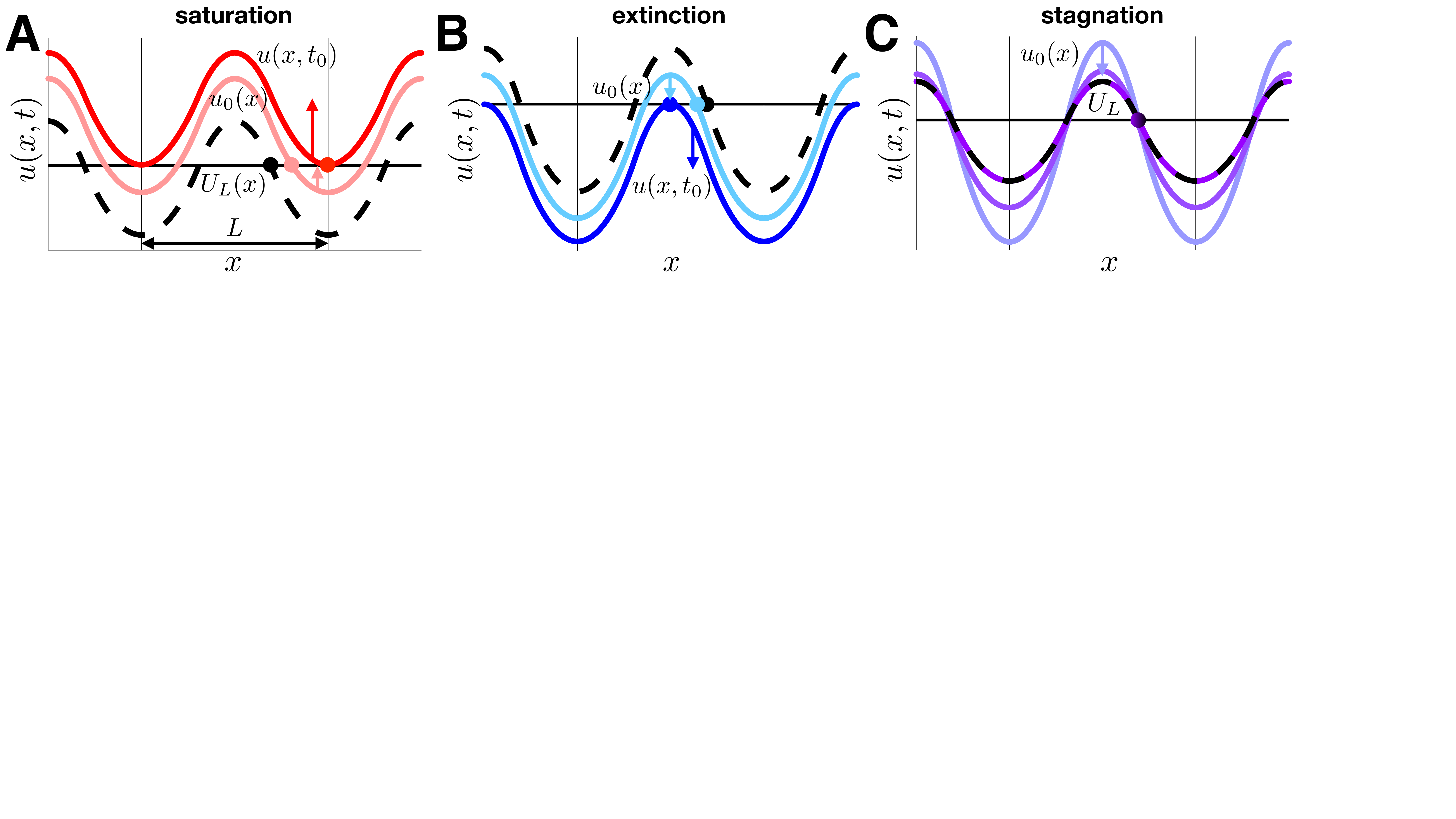} \end{center}
\caption{Long term behavior of $u(x,t)$, given a periodic initial condition with $u_L(\pm \ell_L + nL) = \kappa$, $\forall n \in \Z$, depends on relation between $\ell_L$ and $b_L = {\bf W}_L^{-1}(\kappa)$. (A) If $\ell_L > b_L$, saturation occurs, and $\lim_{t \to \infty} u(x,t) \equiv 1$. (B) If $\ell_L< b_L$, extinction occurs, and $\lim_{t \to \infty} u(x,t) \equiv 0$. (C) If $\ell_L = b_L$, stagnation occurs, and $\lim_{t \to \infty} u(x,t) = U_L(x)$ as in Eq.~(\ref{statper}).}
\label{fig6_periodic}
\end{figure}

\noindent
{\bf Saturation.} If $\ell_L > {\bf W}_L^{-1}(\kappa)$, then $a'(t)>0$ and $\alpha (t) < 0$ for a time interval $t \in (0,t_0)$. At $t_0>0$, $a(t_0) = L/2$, and the interface dynamics, Eq.~(\ref{periface}), breaks down. We can be sure this occurs in finite time because ${\bf W}_L(a(t)) - \kappa$ is positive and increasing in $a(t)$. In addition $\lim_{t \to t_0^-} \alpha (t_0) = 0$, since $u_0'(L/2 + nL)=0$, $\forall n \in \Z$. Thus, at $t_0< \infty$, we have $u(x,t_0) \geq \kappa$, and $\lim_{t \to \infty} u(x,t) = 1$, due to our analysis in Section \ref{homstates} (See Fig. \ref{fig6_periodic}A). \\
\vspace{-3mm}

\noindent
{\bf Extinction.} If $\ell_L < {\bf W}_L^{-1}(\kappa)$, then $a'(t)< 0$ and $0< a(t) < \ell_L$ on $t \in (0,t_0)$. At finite $t_0>0$, $a(t_0) = 0$ and $\alpha (t_0) = 0$, so Eq.~(\ref{periface}) break down, as before. Subsequently, $u(x,t_0) \leq \kappa$, so $\lim_{t \to \infty} u(x,t) =0$, as in Section \ref{ifacecrit} (See Fig. \ref{fig6_periodic}B). \\
\vspace{-3mm}

\noindent
{\bf Stagnation.} If $\ell_L = {\bf W}_L^{-1}(\kappa)$, then $a'(t) =0$ assuming $\alpha (t) <0$, implying $a(t) \equiv \ell_L$ for $t>0$. Plugging into Eq.~(\ref{periface}) yields $\alpha (t) = -|U_L'(b_L)| (1-\e^{-t}) + u_0'(b_L) \e^{-t} <0$ for $t>0$, for $|U_L'(b)|$ defined in Eq.~(\ref{uperprime}). Thus, we can explicitly solve for
\begin{align*}
u(x,t) = U_L(x) + \e^{-t} \left[ u_L(x) - U_L(x)  \right],
\end{align*}
so $\lim_{t \to \infty} U_L(x)$, as defined in Eq.~(\ref{statper}) (See Fig. \ref{fig6_periodic}C). \\
\vspace{-2mm}

\noindent
{\bf Asymptotic results.} Similar to the single active region case, we can obtain leading order approximations for the transient dynamics approaching the homogeneous states. For periodic initial conditions, we do not obtain traveling waves in the long time limit. In the case of saturation, we can estimate the time $t_0$ at which $u(x,t_0) \geq \kappa$, assuming $L/2 -a(t)$, $\alpha (t)$, and $t_0$ are small. We approximate $\alpha (t) \approx u_L''(L/2)(\ell_L - L/2)$, so $t_0 \approx (L-2\ell_L)^2 u_L''(L/2)/[2 - 4\kappa]$. \\[0.5ex] In the case of extinction, the calculation is quite similar to that presented in Section \ref{asymexp}, and we find $u(x,t_0) \leq \kappa$ at $t_0 \approx \ell_L^2 |u_L''(0)|/ \kappa$ in the limit $0<\ell_L\ll 1$. \\
\vspace{-2mm}

\begin{figure}
\begin{center} \includegraphics[width=10cm]{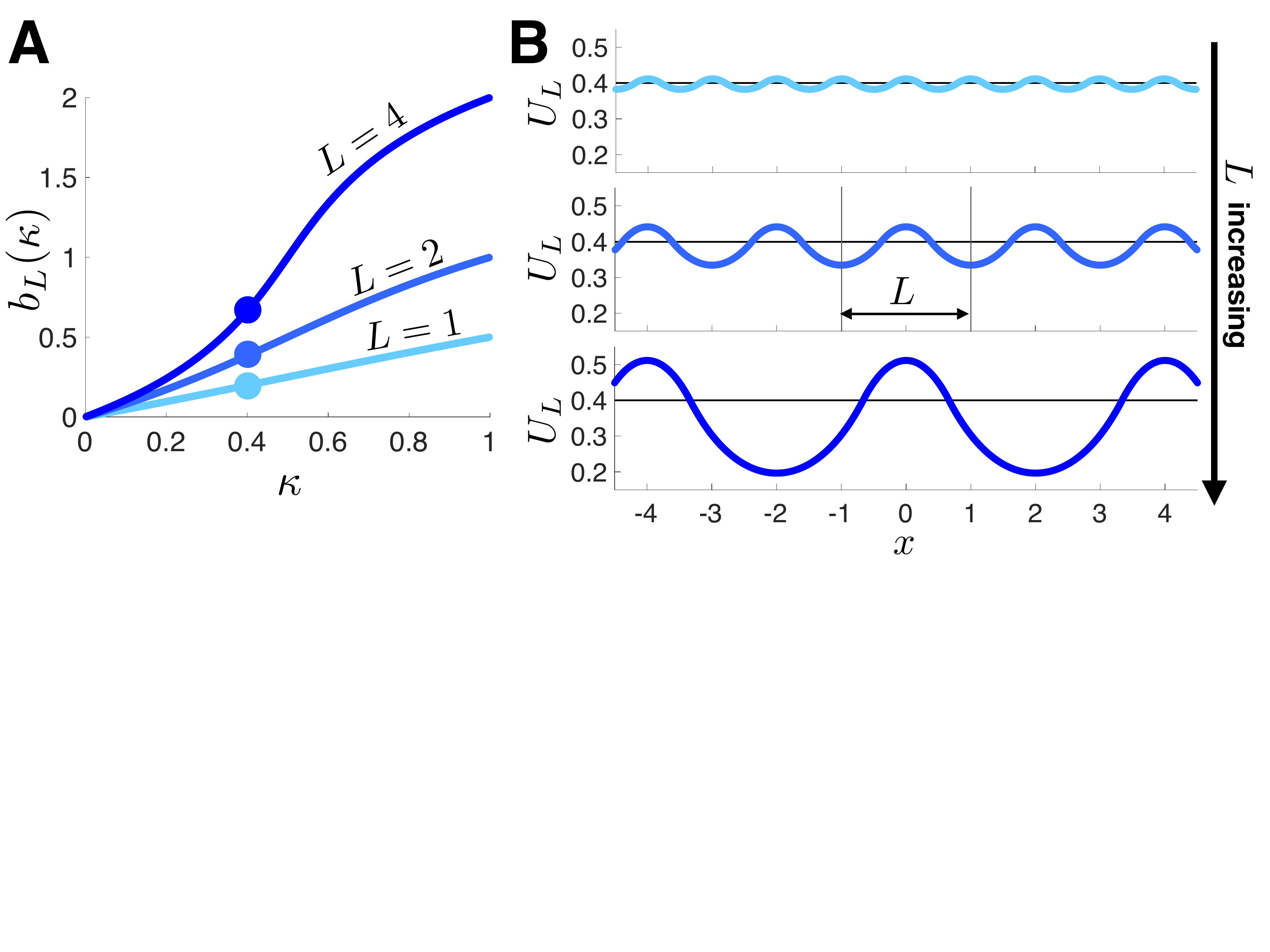} \end{center}
\caption{Stationary periodic solutions $U_L$ to Eq.~(\ref{nfield}) with exponential kernel, Eq.~(\ref{exp}). (A) Half-width $b_L$ of the active region on each period $L$ increases with threshold $\kappa$ and period $L$, as determined by Eq.~(\ref{expper}). (B) Fixing $\kappa = 0.4$ and increasing $L$ yields periodic patterns (corresponding to dots in panel A) with wider active regions and higher amplitude oscillations, as determined by Eq.~(\ref{persolexp}).}
\label{fig7_perexp}
\end{figure}

\noindent
{\bf Exponential kernels.} Assuming $w(x)$ is given by Eq.~(\ref{exp}), we can obtain a simple implicit expression for the critical half-width $b_L : = {\bf W}_L^{-1}(\kappa)$. Plugging Eq.~(\ref{exp}) into Eq.~(\ref{WL}), we can write the threshold condition $U_L(\pm b + nL) = \kappa$ in the following form
\begin{equation*}
\kappa = \frac{1 - \e^{-2b}}{2} \left[ 1 +  \sum_{n =1}^{\infty} \e^{-nL} + \e^{2b} \sum_{n=1}^{\infty} \e^{-nL} \right] = \frac{1 - \e^{-2b}}{2(1+ \e^{-L})} \left[ 1 + \e^{2b - L} \right],  
\end{equation*}
which simplifies to 
\begin{equation}
\kappa = \frac{\sinh (b)}{\sinh (L/2)} \cosh (L/2-b) : = {\bf W}_L(b).  \label{expper}
\end{equation}
Clearly, ${\bf W}_L(0) = 0$ and ${\bf W}_L(L/2) = \cosh (0) = 1$, and ${\bf W}_L'(b) = \cosh (b) \cosh (L/2-b) - \sinh(b) \sinh(L/2-b) >0$ because $\cosh(x) > \sinh (x)$, $\forall x \in \R$. Thus, for any $\kappa \in (0,1)$, Eq.~(\ref{expper}) will have a solution, as expected. Eq.~(\ref{expper}) must be solved numerically (Fig. \ref{fig7_perexp}A), showing $b_L$ increases with $\kappa$ and $L$. Lastly, we can apply a similar reduction to the resulting formula for the solution $U_L$ to yield
\begin{align}
U_L(x) = \left\{ \begin{array}{cc} \frac{\D \sinh (b)}{\D \sinh(L/2)} \cosh \left( \frac{\D L}{\D 2} + x_n \right) , & x_n \in (b-L,-b), \\
1 - \frac{\D \e^{L-b} - \e^b}{\D \e^L-1} \cosh(x_n), & x_n \in (-b,b), \\
\frac{\D \sinh (b)}{\D \sinh(L/2)} \cosh \left( \frac{\D L}{\D 2} - x_n \right) , & x_n \in (b,L-b), \end{array} \right.  \label{persolexp}
\end{align}
where we define $x_n : = x-nL$, $ \forall n \in \Z$ (Fig. \ref{fig7_perexp}B). Note, we obtain the threshold condition, Eq.~(\ref{expper}) for $U_L(\pm b + nL)$, $ \forall n \in \Z$, and as $L \to \infty$, $U_L(x) \to U_b(x)$~\cite{kilpatrick10}.

\subsection{Two symmetric active regions} We now consider the case of two symmetric active regions in the initial conditions. More specifically, we consider a class of bimodal even initial conditions $u_0(x)=u_0(-x)$, with two active regions supported in $[-\ell_2,-\ell_1]\cup[\ell_1,\ell_2]$ for $0<\ell_1<\ell_2$. That is, we have $u_0(x)\geq \kappa $ for all $x\in [-\ell_2,-\ell_1]\cup[\ell_1,\ell_2]$ and $u_0(x)<\kappa$ elsewhere, with $u_0'(x)\gtrless 0$ for $x \lessgtr \mp \ell_2$. We also ensure a non-degeneracy condition of the derivative of $u_0$ at the boundaries of the active regions, namely $u_0'(\pm\ell_{1,2})\neq0$.  These hypotheses on the initial conditions ensure that, as time evolves, the active regions can be described by $a(t) : = a_2(t) = -b_1(t)$, $b(t): = b_2(t) = - a_1(t)$, $\alpha (t) := \alpha_2(t) = - \beta_1(t)$, and $\beta (t) : = \beta_2(t) = - \alpha_1(t)$. We can therefore write the system of interface equations and their gradients, Eq.~(\ref{multifacesys}) in the following simpler form
\begin{subequations} \label{twosymiface}
\begin{align}
a'(t) &= - \frac{1}{\alpha (t)} \left[ W(b(t) - a(t)) - \kappa + W(b(t) + a(t)) - W(2a(t)) \right], \\
b'(t) &= - \frac{1}{\beta (t)} \left[ W(b(t) - a(t)) - \kappa + W(2b(t)) - W(b(t) + a(t)) \right], \\
\alpha (t) &= u_0'(a(t)) \e^{-t} + \e^{-t} \int_0^t \e^s \left[ w(a(t)+b(s)) - w(a(t)+a(s))  \right] \d s , \nonumber \\
&~~~+ \e^{-t} \int_0^t \e^s \left[ w(a(t)-a(s)) - w(a(t)-b(s)) \right] \d s,\\ 
\beta (t) &=  u_0'(b(t)) \e^{-t} +\e^{-t} \int_0^t \e^s \left[ w(b(t)-a(s)) - w(b(t) - b(s)) \right] \d s\nonumber\\
 &~~~+\e^{-t} \int_0^t \e^s \left[ w(b(t)+b(s)) - w(b(t) +a(s)) \right] \d s
\end{align}
\end{subequations}
The system Eq.~(\ref{twosymiface}) is closed by the initial conditions $a(0) = \ell_1$ and $b(0) =\ell_2$. As opposed to the single active region case, it is not possible to develop a simple condition on $(\ell_1,\ell_2)$ that determines whether propagation, extinction, or stagnation occurs in the long time limit. However, we can still partition the space of initial conditions $(\ell_1,\ell_2)$ into several cases, for which the long term behavior of Eq.~(\ref{nfield}) is determined by the initial transient dynamics of $(a(t),b(t))$. The first simple observation that can be made is that both $W(b(t) + a(t)) - W(2a(t)) > 0$ and $W(2b(t)) - W(b(t) + a(t))> 0$ for all time whenever they are well defined ({\it i.e.} as long as $0<a(t)<b(t)$). As a consequence, we can already rule out the trivial case where $\ell_2-\ell_1\geq W^{-1}(\kappa)$.  \\
\vspace{-3mm}

\noindent
{\bf Class I: $\ell_2-\ell_1 \geq W^{-1}(\kappa)$.} In this case, we automatically deduce that $b'(t)>0$ while $a'(t)<0$ for all time where they are both well defined. This implies that there exists a finite $t_*>0$ at which we have $a(t_*)=0$. At this point, the two active regions merge to form a single active region given at time $t=t_*$ by $[-b(t_*),b(t_*)]$ with $2b(t_*)>2\ell_2>W^{-1}(\kappa)$ as  $\ell_2-\ell_1 \geq W^{-1}(\kappa)$. As a consequence, we are back to the propagation scenario studied in Section~\ref{ifacecrit} and we find the associated solution of the neural field Eq.~\eqref{nfield} verify $u\rightarrow 1$ locally uniformly on $x \in \R$ as $t\rightarrow+\infty$. \\
\vspace{-3mm}

\begin{figure}
 \begin{center}\includegraphics[width=13cm]{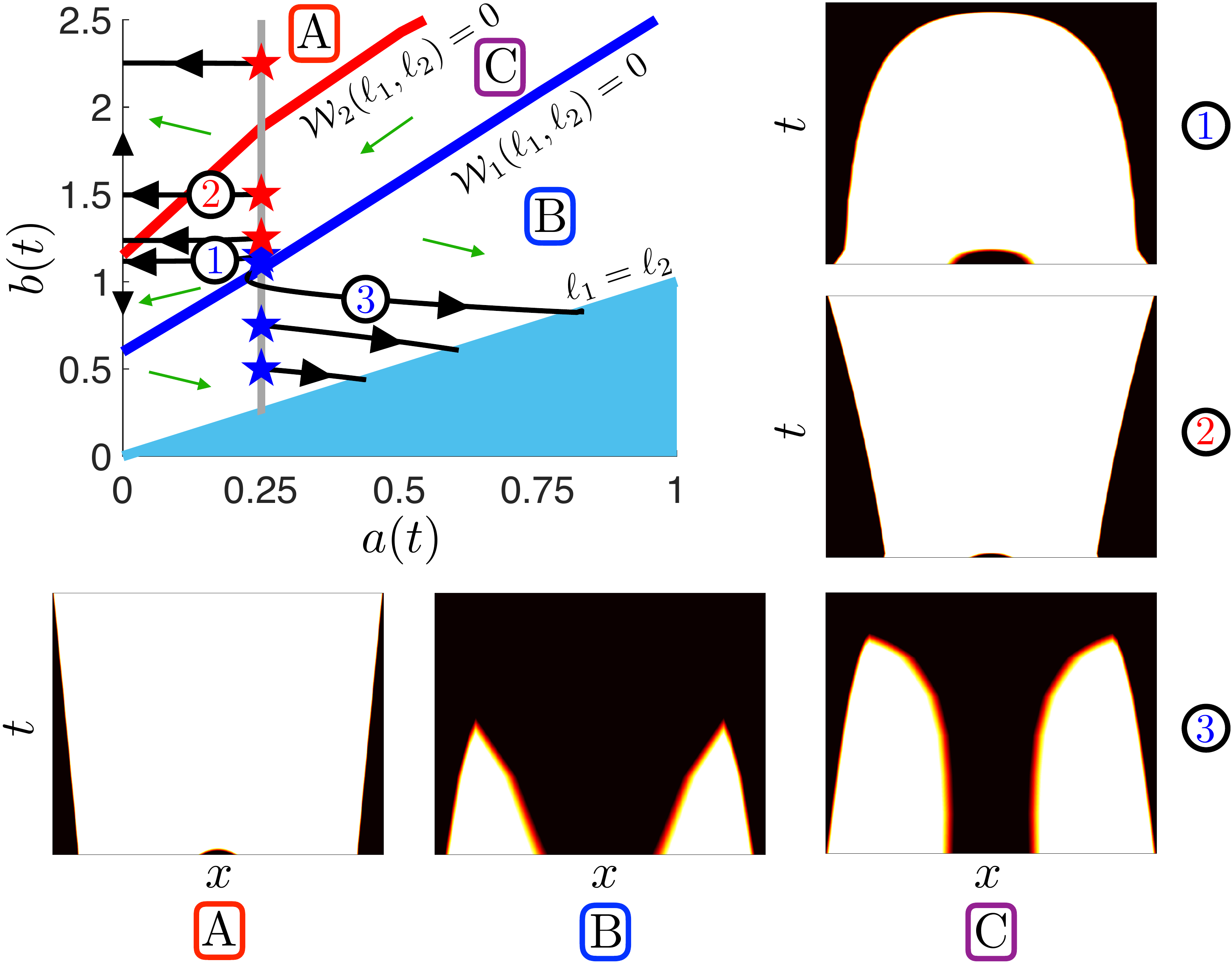} \end{center}
 \vspace{-2mm}
 \caption{Evolution of interfaces for two symmetric bumps having Class II initial conditions: $\ell_2-\ell_1<W^{-1}(\kappa)$. Phase portrait of Eq.~\eqref{twosymiface} in ($a(t)$,$b(t)$) is shown in the case of an exponential kernel, Eq.~(\ref{exp}), $\kappa=0.45$ and the initial condition $u_0(x)$, Eq.~\ref{eqICd}. We fix $a(0)=\ell_1=1/4$ and vary $b(0) = \ell_2$ from $0.5$ to $2.25$. Some initial conditions (red stars) lead to trajectories (black lines) that propagate, while other initial conditions (blue stars) lead to extinction. Case A ($\W_1(\ell_1,\ell_2)>\W_2(\ell_1,\ell_2)\geq 0$): $a(t)$ vanishes in finite time with a final value above the nullcline $\W_2(\ell_1,\ell_2)=0$ (where red line meets $\ell_2$ axis), leading to propagation. See corresponding example evolution of $u(x,t)$. Case B ($0\geq \W_1(\ell_1,\ell_2)>\W_2(\ell_1,\ell_2)$): $a(t)$ and $b(t)$ merge in finite time, leading to extinction. Case C ($\W_1(\ell_1,\ell_2)>0>\W_2(\ell_1,\ell_2)$): Three subcases are described in main text, leading to either extinction for subcases (1) and (3) or propagation for subcase (2). Green arrows indicate the direction of the vector field in each sub-region. Outer panels demonstrate behavior of the full neural field model, Eq.~(\ref{nfield}), in the cases A, B, C1, C2, and C3.}
 \vspace{-4mm}
 \label{fig8_region}
 \end{figure}

\noindent
{\bf Class II: $\ell_2-\ell_1 < W^{-1}(\kappa)$.} We now discuss the case where $\ell_2-\ell_1< W^{-1}(\kappa)$. In order to simplify the presentation, we define the following two quantities:
\begin{align*}
\W_1(\ell_1,\ell_2)&:=W(\ell_2-\ell_1)-\kappa + W(\ell_1+\ell_2)-W(2\ell_1),\\
\W_2(\ell_1,\ell_2)&:=W(\ell_2-\ell_1)-\kappa + W(2\ell_2)-W(\ell_1+\ell_2),
\end{align*}
defined for all $0<\ell_1<\ell_2$. It is crucial to observe that $\W_1(\ell_1,\ell_2)-\W_2(\ell_1,\ell_2)=2W(\ell_1+\ell_2)-W(2\ell_1)-W(2\ell_2)>0$ for any $0<\ell_1<\ell_2$ by concavity of the function $W$ on the positive half line. Thus, we only have to consider three cases (See Fig. \ref{fig8_region}).

{\em Case A:} If $\W_1(\ell_1,\ell_2)>\W_2(\ell_1,\ell_2)\geq 0$, then $b'(t)>0$ and $a'(t)<0$ for all time where they are both well defined. Once again, there must exist $t_*>0$ at which $a(t_*)=0$. At that point, the two active regions merge to form a single active region given at time $t=t_*$ by $[-b(t_*),b(t_*)]$ with $2b(t_*)>2\ell_2>W^{-1}(\kappa)$. Indeed, from $\W_2(\ell_1,\ell_2)>0$, we deduce that $W(2\ell_2)>W(\ell_1+\ell_2)-W(\ell_2-\ell_1)+\kappa>\kappa$. And we are back to the propagation case of Section~\ref{ifacecrit}.

{\em Case B:} If $0\geq \W_1(\ell_1,\ell_2)>\W_2(\ell_1,\ell_2)$, then $b'(t)<0$ and $a'(t)>0$ for all time where they are both well defined. As a consequence, there exists some time $t_{*}>0$ where $a(t_{*})=b(t_{*})$ and such that $u(x,t_{*})\leq \kappa$ for all $x\in\R$. As a consequence, this will lead to the extinction case of Section~\ref{ifacecrit} and we get that the solutions of the neural field equation Eq.~\eqref{nfield} verify $u\rightarrow 0$ uniformly on $\R$ as $t\rightarrow+\infty$.

{\em Case C:}  If $\W_1(\ell_1,\ell_2)>0>\W_2(\ell_1,\ell_2)$, then we are led to study three sub-cases:

{\it Sub-case 1:} Both $a(t)$ and $b(t)$ satisfy $\W_1(a(t),b(t))>0>\W_2(a(t),b(t))$ for all $t\in[0,t_*)$ where they are well defined, and at time $t=t_*$ we have $a(t_*)=0$. Once more, at this point, the two active regions merge to form a single active region at time $t=t_*$: $[-b(t_*),b(t_*)]$ with $2b(t_*)<2\ell_2$. Thus, it is enough to check that the limit $t\rightarrow t_*$, we have $\W_2(a(t),b(t))\rightarrow W(2 b (t_*))-\kappa$. Since $\W_2$ does not change sign in $(0,t_*)$, then $0\geq W(2 b (t_*))-\kappa$, so we obtain either stagnation (when $W(2 b (t_*))=\kappa$) or extinction (when $W(2 b (t_*))<\kappa$), as studied in Section~\ref{ifacecrit}. 

 {\it Sub-case 2:} There exists a time $t_0>0$ where $a(t)$ and $b(t)$ satisfy $\W_1(a(t),b(t))>0>\W_2(a(t),b(t))$ for all $t\in[0,t_0)$, and at $t=t_0$ we have $a(t_0)\neq0$ with $\W_2(a(t_0),b(t_0))=0$ while $\W_1(a(t_0),b(t_0))>0$, in which case we are back to Case A and propagation occurs.
 
 {\it Sub-case 3:} There exists a time $t_1>0$ where $a(t)$ and $b(t)$ satisfy $\W_1(a(t),b(t))>0>\W_2(a(t),b(t))$ for all $t\in[0,t_1)$, and at $t=t_1$ we have $a(t_1)\neq0$ with  $\W_1(a(t_1),b(t_1))=0$ while $0>\W_2(a(t_1),b(t_1))$, in that case we are back to Case B and extinction occurs.
 
We illustrate these different scenarios on a specific example in Fig. \ref{fig8_region} using an exponential kernel, Eq.~(\ref{exp}), and the following initial condition
\begin{equation}
u_0(x) = \frac{U_0}{2}\left( e^{-|x+x_0|}+e^{-|x-x_0|} \right),
\label{eqICd}
\end{equation}
which allows us to specify
 \begin{equation*}
 x_0 = \frac{1}{2}\ln \left(-1+2\cosh(\ell_1)e^{\ell_2} \right), \text{ and } U_0=\kappa \frac{\sqrt{-1+2\cosh(\ell_1)e^{\ell_2}}}{\cosh(\ell_1)},
 \end{equation*}
and ensure that $u_0(\pm\ell_{1,2})=\kappa$. Note, for a fixed $\ell_1$, there is a critical value of $\ell_2$ at which initial conditions transition from leading to extinction (blue stars) to those that lead to propagation (red stars) in Fig. \ref{fig8_region}. Corresponding example simulations of the full neural field Eq.~(\ref{nfield}) are also shown.

\subsection{Critical spatially-periodic stimuli} Finally, we can consider the impact of spatially periodic inputs $I(x,t) = I(x) \chi_{[0,t_1]}$ ($I(x) = I(x+L)$) on the long-term dynamics of Eq.~(\ref{nfield}), assuming $u_0(x) \equiv 0$. To make our calculations more straightforward, we assume that $I(x)$ is even and unimodal on $x \in [-L/2,L/2]$ with $I'(0) = I'(\pm L/2) = 0$. Our analysis follows similar principles as that performed for unimodal inputs in Section \ref{critstim}. To ensure propagation, there must be no stationary bump solutions to Eq.~(\ref{nfield}) with stationary input $I(x)$ and the active region on $x \in [-L/2,L/2]$ at $t=t_1$ must be wider than $b_L = {\bf W}_L^{-1}(\kappa)$.

\begin{figure}
 \begin{center} \includegraphics[width=13cm]{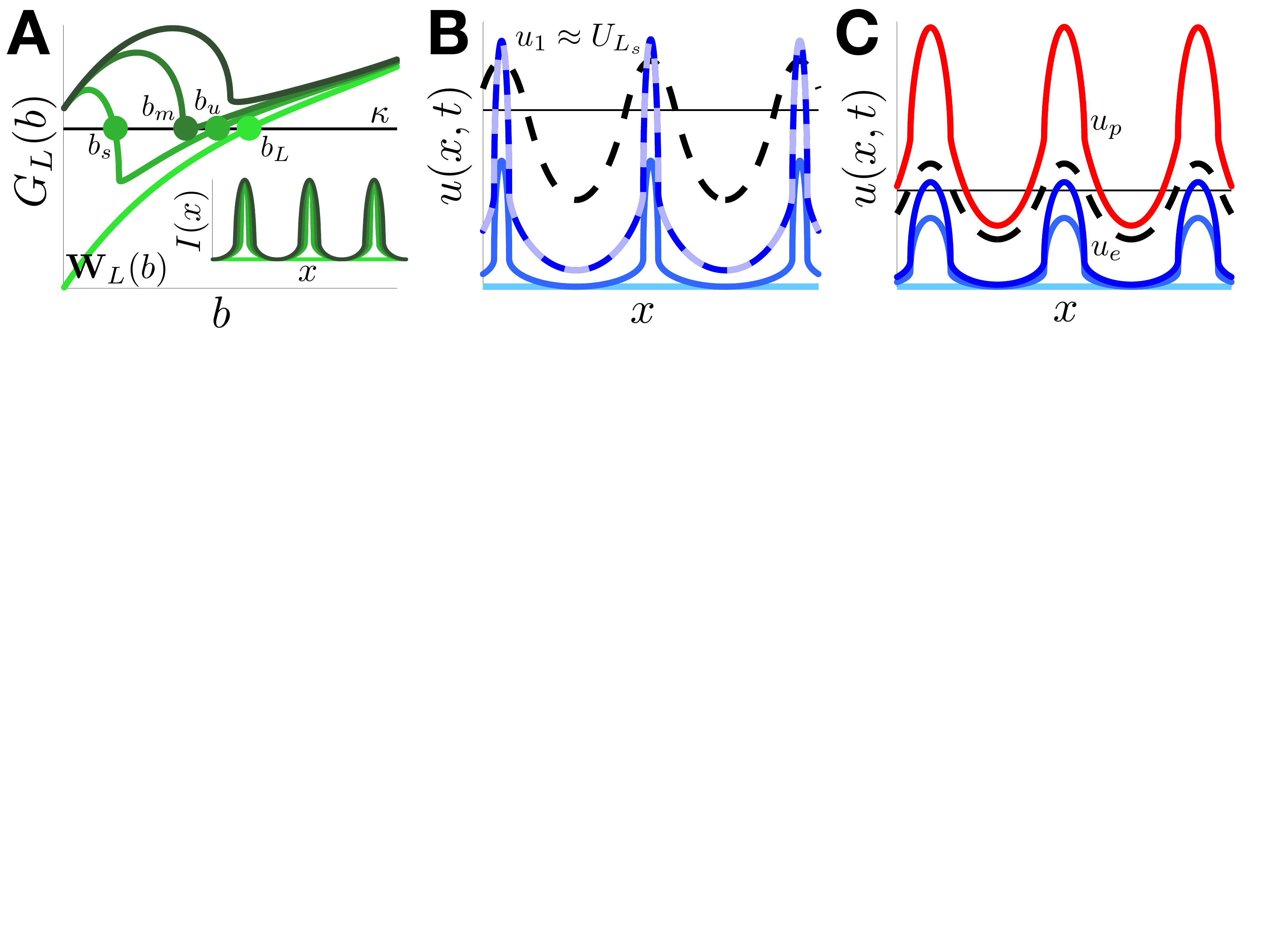} \end{center}
 \vspace{-2mm}
 \caption{Conditions for propagation driven by a spatially-periodic input $I(x,t) = I(x) \chi_{[0,t_1]}$ with $I(x) = I(x+L)$, and $u(x,0) \equiv 0$. (A) For periodic, even, and positive profile $I(x)$ with $I(0)>\kappa$, propagation only occurs if $G_L(b) = {\bf W}_L(b) + I(b) = \kappa$ has no solutions. If solutions to Eq.~(\ref{perinpwid}) exist, the minimal one is linearly ($b_s$) or marginally ($b_m$) stable. For inputs $I(x)$ that are monotone decreasing on $x \in (0,L/2)$, there are only two solutions. (B) If $G_L(b_s) = \kappa$ is satisfied for some $b_s$, $u(x,t_1) \approx U_{L_s}(x)$ for large $t_1$ with active region centered at $x=0$ given $[-a_s,a_s]$ where $a_s < b_L$, so $\lim_{t \to \infty} u(x,t) \equiv 0$. (C) Here $I(x)$ is chosen so that $G_L(b) = \kappa$ has no solutions. If $t_1 := t_e$, then $u_e(\pm a_e) := u(\pm a_e,t_e) = \kappa$ and $a_e < b_L$, so $\lim_{t \to \infty} u(x,t) \equiv 0$. However, for $t_1 := t_p$ with $u_p(\pm a_p) : = u(\pm a_p,t_e) = \kappa$ and $a_p> b_L$, then $\lim_{t \to \infty} u(x,t) \equiv 1$.}
 \vspace{-4mm}
 \label{fig9_perinp}
 \end{figure}

Stationary periodic patterns exist as solutions to Eq.~(\ref{nfield}) for $I(x,t) = I(x)$ periodic ($I(x) = I(x+L)$), even, and unimodal on $x \in [-L/2,L/2]$. Adapting our analysis from Section \ref{persolns}, we can show they have the form
\begin{align*}
U_L(x) = \sum_{n \in \Z} (W(x+b+nL)-W(x-b+nL)) + I(x).
\end{align*}
Applying the threshold conditions, $U_L(\pm b + nL) = \kappa$ then yields
\begin{align}
\sum_{n \in \Z} (W(2b+nL)-W(nL)) + I(b) = {\bf W}_L(b) + I(b) =  G_L(b) = \kappa, \label{perinpwid}
\end{align}
which defines an implicit equation for the half-width $b$ of each active region. Any solutions to Eq.~(\ref{perinpwid}) will be less than the solutions to the input-free case $I \equiv 0$: $b < b_L = {\bf W}_L^{-1}(\kappa)$, as in Eq.~(\ref{WL}). Subtracting Eq.~(\ref{perinpwid}) from Eq.~(\ref{WL}), we find ${\bf W}_L(b_L) - {\bf W}_L(b) = I(b) >0$, so ${\bf W}_L(b_L) > {\bf W}_L(b)$, so $b_L>b$, since ${\bf W}_L(x)$ is monotone increasing (Fig. \ref{fig9_perinp}A). Local stability with respect to $L$-periodic perturbations can be characterized by deriving a linearized equation for $\psi (x)$, where $u(x,t)  = U_L(x) + \e^{\lambda t} \psi (x)$. Repeating the argument of Section~\ref{persolns}, the most unstable part of the spectrum is given by  even $L$-periodic perturbations, $\psi (\pm b+nL) = \psi_e \neq 0$, with associated eigenvalue
\begin{align*}
\lambda_e = \frac{2 \sum_{n \in \Z} w(2b+nL) + I'(b)}{\sum_{n \in \Z} (w(nL)-w(2b+nL)) - I'(b)} = \frac{G'(b)}{\sum_{n \in \Z} (w(nL)-w(2b+nL)) - I'(b)},
\end{align*}
so the sign of $\lambda_e$ is the same as the sign of $G'(b)$. Thus, the bump is stable with respect to $L$-periodic perturbations if $G'(b) <0$. When $I(0) > \kappa$, $G(0)> \kappa$, and since $G(x)$ is continuous, if there are any solutions to Eq.~(\ref{perinpwid}), the minimal one will be stable or marginally stable, since $G(x)$ will be decreasing or at a local minimum.

We now demonstrate that for a spatiotemporal input, $I(x,t) = I(x) \chi_{[0,t_1]}$, to generate a saturating solution, (i) Eq.~(\ref{perinpwid}) must have no solutions and (ii) $t_1$ must be large enough so the active region $A(t) = \cup_{n \in \Z} [-a(t) + nL, a(t)+nL ]$ satisfies $a(t_1) > b_L$, where $b_L$ solves Eq.~(\ref{perinpwid}) for $I \equiv 0$. Starting from $u_0(x) \equiv 0$, we know initially the dynamics obeys $\pd_tu(x,t) = -u(x,t) + I(x,t)$, so
the time needed to produce a nontrivial active region is given by $t_0 = \ln \left[ \frac{I(0)}{I(0) - \kappa} \right]$, as before. If $t_1 \leq t_0$, then the long term dynamics of the solution is $u(x,t) = I(x) (1-\e^{-t_1}) \e^{-(t-t_1)}$ for $t > t_1$, so $\lim_{t \to \infty} u(x,t) \equiv 0$. Note if $I(0)< \kappa$, then $u(x,t) < \kappa$ clearly for all $t>0$. 

If $t_1 > t_0$, then for $t_0 < t < t_1$, we can derive the interface equations for $u(\pm a(t), t) = \kappa$, similar to Eq.~(\ref{periface}), finding
\begin{subequations} \label{perinpint}
\begin{align}
a'(t) &= - \frac{1}{\alpha(t)} \left[ {\bf W}_L(a(t)) - \kappa + I(a(t)) \right], \\
\alpha (t) &= u_L'(a(t)) \e^{-t} + \e^{-t} \int_0^t \e^{s} \left[ \sum_{n \in \Z} w_n(a(t),a(s)) + I(a(s)) \right] \d s,
\end{align}
\end{subequations}
with initial conditions $a(t_0) = 0$ and $\alpha (t_0) = 0$, so $a'(t_0)$ diverges. As before, we can desingularize Eq.~(\ref{perinpint}) with the change of variables $\tau = - \int_{t_0}^t \frac{\d s}{\alpha (s)}$, so we can write a differential equation for $\tilde{a}(\tau)$ in $\tau$ as
\begin{align}
\frac{\d \tilde{a}}{\d \tau} = {\bf W}_L(\tilde{a}(\tau)) - \kappa + I(\tilde{a}(\tau)),  \label{atilper}
\end{align}
with $\tilde{a} (0) = 0$. Since $\alpha (t)<0$ for $t>t_0$, $\tau$ will be an increasing function of $t$, so we now refer to $\tau_1 : = \tau (t_1)$ and note $\tau (t_0) = 0$. By assumption $I(0) - \kappa > 0$, so $\frac{\d \tilde{a}}{\d \tau} > 0$ for all $\tau$ where it is defined.

We now discuss the three remaining cases: (I) Eq.~(\ref{perinpwid}) has at least one solution, so saturation does not occur; (II) Eq.~(\ref{perinpwid}) has no solutions, but $\tau_1 \leq \tau_c$, the time at which $\tilde{a}(\tau_c) = b_L$ for $I(x,\tau) \equiv I(x)$, and saturation does not occur; (III) Eq.~(\ref{perinpwid}) has no solutions, and $\tau_1 > \tau_c$, so saturation occurs.  \\
\vspace{-3mm}

\noindent
{\em Case I:} $\min_{x \in \R} G(x) \leq \kappa$. Here, Eq.~(\ref{perinpwid}) has at least one solution. Since we have assumed $I(0)> \kappa$, this solution $b_{\rm min}$ is linearly or marginally stable with respect to even and odd perturbations. Eq.~(\ref{atilper}) implies $\frac{\d \tilde{a}}{\d \tau} > 0$ for all $ \tau < \tau_1$, but $\frac{\d \tilde{a}}{\d \tau}$ vanishes at $\tilde{a} = b_{\rm min}$, so $\tilde{a}(\tau) < b_{\rm min} < b_0$ for all $\tau < \tau_1$. Thus, once $\tau = \tau_1$, the dynamics is described by the extinction case from Section \ref{peric}, and $\lim_{t \to \infty} u(x,t) \equiv 0$ (Fig. \ref{fig9_perinp}B). \\
\vspace{-3mm}

\noindent
{\em Case II:} $\min_{x \in \R} G(x) > \kappa$ {\em and} $\tau_1 \leq \tau_c$. Here Eq.~(\ref{perinpwid}) has no solutions, but $\tilde{a}(\tau)$ will not grow large enough for saturation to occur once $I(x,\tau) = 0$, since $\tau_1 \leq \tau_c$. We define $\tau_c$ as the critical time when $\tilde{a}(\tau_c) = b_L = {\bf W}_L^{-1} (\kappa)$, given by the formula
\begin{align}
\int_0^{{\bf W}_L^{-1}(\kappa)} \frac{\d a}{W_L(a) - \kappa + I(a)} = - \int_{t_0}^{t_c} \frac{\d t}{\alpha (t)} : = \tau_c.  \label{pertauc}
\end{align}
By definition, $\tilde{a}(\tau_1) \leq b_L$, so once $\tau = \tau_1$, the dynamics is described by either (a) the extinction case in Section \ref{peric} if $\tau_1 < \tau_c$, so $\lim_{t \to \infty} u(x,t) \equiv 0$, or (b) the stagnation case in Section \ref{peric} if $\tau_1 = \tau_c$, so $\lim_{t \to \infty} u(x,t) \equiv U_L(x)$ (Fig. \ref{fig9_perinp}C). \\
\vspace{-3mm}

\noindent
{\em Case III:} $\min_{x \in \R} G(x) > \kappa$ {\em and} $\tau_1> \tau_c$. Requiring $\tau_1 > \tau_c$ with Eq.~(\ref{pertauc}), we have that $\tilde{a}(\tau_1) > b_L$. Thus, after $\tau = \tau_1$, the dynamics is described by the saturation case in Section \ref{peric}, so $\lim_{t \to \infty} u(x,t) \equiv 1$ (Fig. \ref{fig9_perinp}C).

\section{Discussion}
\label{discussion}
In this paper, we have studied threshold phenomena of front propagation in the excitatory neural field Eq.~(\ref{nfield}) using an interface dynamics approach. Our interface analysis projects the dynamics of the integrodifferential equations to a set of differential equations for the boundaries of the active regions, where the neural activity is superthreshold. The interface equations can be used to categorize initial conditions or external stimuli based on whether the corresponding long term dynamics of the neural field are extinction ($u \to 0$), propagation/saturation ($u \to 1$), or stagnation ($u \to U_{\rm stat}(x) \not\equiv 0,1$). We considered several classes of initial conditions, which admit explicit results: (i) functions with a single active region; (ii) even and periodic functions with an infinite number of active regions; and (iii) a two-parameter family of even functions with two active regions.
In these particular cases, the conditions for extinction, propagation/saturation, or stagnation can be expressed in terms of a few inequalities for the parameters specifying the initial conditions.
We were able to obtain a similar trichotomy when the neural field Eq.~(\ref{nfield}) is forced by a fixed critical stimulus (e.g., unimodal and periodic) over a finite time interval. Our analysis assumes the nonlinearity in the neural field arises from a Heaviside firing rate function, so the dynamics of the neural field Eq.~(\ref{nfield}) can be equivalently expressed as differential equations for the spatial locations where the neural activity equals the threshold of the firing rate function. This work addresses an important problem in the analysis of models of large-scale neural activity, determining the long term behavior of neuronal network dynamics that begin away from equilibrium.


There are several natural extensions of this work that build on the idea of developing critical thresholds for propagation in neural fields using an interface dynamics approach. For instance, one possibility would be to consider a planar version of Eq.~(\ref{nfield}), and develop closed form equations for the corresponding interface dynamics of the contours encompassing active regions as in \cite{coombes12}. In a preliminary analysis, we have already found that our results developed herein for single active regions can be extended to the case of radially symmetric initial conditions in two-dimensions (2D). Single stripe and periodic stripe patterns may also admit explicit analysis. However, there are also a number of other classes of initial condition that do not have a one-dimensional analogue, which could be interesting to explore, such as spot patterns and multiple concentric annuli. Employing our knowledge of the one-dimensional case may shed light on how to develop a theory for threshold phenomena in 2D. Alternatively, we may also consider neural fields with negative feedback that model adaptation~\cite{pinto01,huang04,kilpatrick10,faye15,avitabile17}, which are known to generate traveling pulses, spiral waves or more exotic phenomena.
In this case, the long term behavior of propagating solutions can be counter-propagating pulses rather than fronts.

\section*{Acknowledgements}

GF received support from the ANR project NONLOCAL ANR-14-CE25-0013. ZPK was supported by an NSF grant (DMS-1615737). ZPK would also like to acknowledge the warm hospitality of the staff and faculty at Institut de Math\'ematiques de Toulouse. This work was partially supported by ANR-11-LABX-0040-CIMI within the program ANR-11-IDEX-0002-02.

\bibliographystyle{siam} 
\bibliography{frontinit}

\end{document}